\pdfoutput=1

\documentclass[useAMS,usenatbib]{mn2e}

\usepackage{graphicx}
\usepackage{hyperref}
\usepackage{amsmath}
\usepackage{relsize}
\usepackage{soul}

\def\Hi{{H\,{\sc\smaller I}}}

\def\Hii{H\,{\sc\smaller II}}
\def\Oii{O\,{\sc\smaller II}}
\def\Nii{N\,{\sc\smaller II}}
\def\Sii{S\,{\sc\smaller II}}

\def\Oiii{O\,{\sc\smaller III}}

\def\aap{A\&A}
\def\aaps{A\&AS}
\def\aj{AJ}
\def\apjs{ApJS}
\def\apj{ApJ}
\def\apjl{ApJL}
\def\apss{Ap\&SS}
\def\araa{ARA\&A}
\def\mnras{MNRAS}
\def\pasp{PASP}
\def\na{New A}

\title[Abundance and kinematic anomalies in HCG~91c]
  {Galaxy Interactions in Compact Groups II: abundance and kinematic anomalies in HCG~91c}

\author[F.P.A.~Vogt et al.]{Fr\'ed\'eric P.A. Vogt$^{1,2}$\thanks{Email: frederic.vogt@anu.edu.au},
   Michael A. Dopita$^{1,3,4}$,
   Sanchayeeta Borthakur$^{2}$,
   \newauthor
   Lourdes Verdes-Montenegro$^{5}$,
   Timothy M. Heckman$^{2}$,
   Min S. Yun$^{6}$,  and
   \newauthor
   Kenneth C. Chambers$^4$\\
   $^1$ Research School of Astronomy and Astrophysics, Australian National University, Canberra, ACT 2611, Australia\\
   $^2$ Department of Physics and Astronomy, Johns Hopkins University, 3400 N. Charles Street, Baltimore, MD 21218, USA\\
   $^3$ Astronomy Department, King Abdulaziz University, P.O. Box 80203, Jeddah, Saudi Arabia\\
   $^4$ Institute for Astronomy, University of Hawaii at Manoa, 2680 Woodlawn Drive, Honolulu, HI 96822, USA\\
   $^5$ Instituto de Astrof\'iõsica de Andaluc\'ia, CSIC, Apdo. Correos 3004, E-18080 Granada, Spain\\
   $^6$ Department of Astronomy, University of Massachusetts, Amherst, MA 01003, USA
   }

\date{Released XXXX Xxxxx XX}

\pagerange{\pageref{firstpage}--\pageref{lastpage}} \pubyear{2015}

\begin{document}

\label{firstpage}

\maketitle

\begin{abstract}

Galaxies in Hickson Compact Group 91 (HCG~91) were observed with the WiFeS integral field spectrograph as part of our ongoing campaign targeting the ionized gas physics and kinematics inside star forming members of compact groups. Here, we report the discovery of {\Hii} regions with abundance and kinematic offsets in the otherwise unremarkable star forming spiral HCG~91c. The optical emission line analysis of this galaxy reveals that at least three {\Hii} regions harbor an oxygen abundance $\sim$0.15 dex lower than expected from their immediate surroundings and from the abundance gradient present in the inner regions of HCG~91c. The same star forming regions are also associated with a small kinematic offset in the form of a lag of 5-10 km s$^{-1}$ with respect to the local circular rotation of the gas. {\Hi} observations of HCG~91 from the \emph{Very Large Array} and broadband optical images from Pan-STARRS suggest that HCG~91c is caught early in its interaction with the other members of HCG~91. We discuss different scenarios to explain the origin of the peculiar star forming regions detected with WiFeS, and show that evidence point towards infalling and collapsing extra-planar gas clouds at the disk-halo interface, possibly as a consequence of long-range gravitational perturbations of HCG~91c from the other group members. As such, HCG~91c provides evidence that some of the perturbations possibly associated with the early phase of galaxy evolution in compact groups impact the star forming disk locally, and on sub-kpc scales.

\end{abstract}


\section{Introduction}\label{sec:intro}

The environment is known to be a key factor influencing the pathways of galaxy evolution. However, the interconnectivity between the large scale, environmental mechanisms (e.g. gravitational interactions, ram pressure stripping) and the internal galactic processes (starburst, star formation quenching, nuclear activity, and both inflows and outflows) remains poorly understood. Part of the issue lies in the fact that galaxies are spatially extended objects, but are often represented with a series of \emph{ensemble properties} in single-spectrum studies. The advent of integral field spectroscopy (IFS) has opened a new way to study galaxies and their evolution on a spatially resolved basis at optical and near-IR wavelengths. The wealth of IFS surveys that have been undertaken in recent years provides direct evidence of the scientific potential of spatially resolved studies of a statistically significant number of galaxies. Such surveys include SAURON \citep[][]{Bacon01, Zeeuw02, Emsellem04}, PINGS \citep[][]{Rosales10}, ATLAS$^{3\rmn{D}}$ \citep[][]{Cappellari11}, CALIFA \citep{Sanchez12},  SAMI \citep{Croom12}, VENGA \citep[][]{Blanc13}, S7 \citep[][]{Dopita15, Dopita14} and MaNGA \citep{Drory14,Bundy14} . 

In this article, we continue our analysis of the ionized gas physics and kinematics associated with star forming galaxies inside compact groups started in \cite{Vogt13}. Compact groups are isolated structures by definition, and are as such often described as \emph{perfect laboratories} to study the consequences of strong, multiple, simultaneous gravitational interactions on galaxies. Compact groups are intrinsically less crowded than clusters \citep[although they can be denser, see][]{Hickson92}, so that the detailed structure of their large scale environment can in principle be better understood. In groups, galaxies are to a first approximation mostly subject to gravitational interactions \citep[][]{Coziol07}, but the role of other processes such as ram-pressure stripping remains uncertain.

Indeed, the presence of a hot halo inside some compact groups has been confirmed observationally with a \emph{Rosat} survey by \cite{Ponman96}. The analysis of \emph{Chandra} observations by \cite{Desjardins13} showed that in some compact groups (and unlike in clusters), the diffuse X-ray emission is associated with individual galaxy members. Yet, other X-ray bright and massive systems have been found to match the X-ray scaling relations of clusters, and may be representative of an evolved state of compact groups \citep[][]{Desjardins14}. Galaxies in compact groups are found to be {\Hi} deficient (when compared to isolated galaxies with similar characteristics). \cite{Verdes-Montenegro01} reported a correlation between {\Hi} deficiency and X-ray emission, suggesting galaxy$\leftrightarrow$hot IGM (Intergalactic Medium) interactions as a possible origin for the observed {\Hi} deficiency. Yet, \cite{Rasmussen08} report in a \emph{Chandra} and \emph{XMM-Newton} study the non-detection of X-ray emission in 4 (out of 8) of the most HI-deficient compact groups of \cite{Verdes-Montenegro01}, suggesting that galaxy $\leftrightarrow$hot IGM interactions (e.g. ram pressure stripping) may in fact not be the  mechanism driving the observed {\Hi} deficiency. The detection of warm H$_{2}$ emission at levels inconsistent with X-ray heating or AGN activity in 32 out of 74 galaxies in 23 compact groups has lead \cite{Cluver13} to propose that shocks induced by galaxies interacting with a cold IGM may be present in many compact groups. The Stefan's Quintet compact group and its intergalactic shock represents an extreme example of this mechanism \citep[][]{Appleton06,Cluver10}.  

Within the larger scheme of galaxy evolution, it has been suggested that galaxies first start to evolve in compact group-like environments, before further processing in clusters. This is usually refereed to as ``pre-processing'' \citep[][]{Cortese06,Vij13}. Pre-processing mechanisms may not be restricted to compact groups and could be active over a wider range of environments \citep[][]{Cybulski14}, but compact groups in particular favor a more rapid evolution of galaxies compared to the field. The existence of a gap in the mid-infrared color distribution for galaxies in compact groups compared to the field has been interpreted as the ability of these dense environments to rapidly transform star forming galaxies into passive ones \citep[][]{Johnson07, Gallagher08, Walker10,Walker12}. Based on the UV properties of galaxies in groups, \cite{Rasmussen12} report that at fixed stellar mass, the specific star formation rate of group members is $\sim$40 per cent less than in the field; a trend best detected for galaxies with masses $\la 10^{9}$ M$_{\odot}$. In the case of elliptical galaxies, \cite{Rosa07} found their stellar population to be older by 1.6 Gyr and more metal-poor by 0.11 dex in [Z/H] compared to similar galaxies in the field, which they interpreted as the signature of truncated star formation in these systems. Yet, most galaxies in compact groups appear to follow the B-band Tully-Fischer relation \citep{Mendes03, Torres10}, as well as the K-band Tully-Fischer relation \citep{Torres13}.

Whether quenched or enhanced, changes in the star formation activity in a galaxy are merely a consequence of the formation, destruction, processing, spatial redistribution and local density variations of the molecular gas (i.e. the fuel) within the system. Therefore, understanding the large scale gas flows in compact groups is key to a better understanding of the evolutionary pathways of galaxies in these environments. For example, \cite{Verdes-Montenegro01} proposed an evolutionary sequence for compact groups, where the {\Hi} gas distribution is first associated with the individual galaxies, before gravitational interactions strip it apart, resulting in no {\Hi} gas left in the galaxies, or possibly (but less frequently) in a large common envelope. 

Here, we describe the use of the WiFeS integral field spectrograph to study the physics and kinematics of the ionized gas in star forming galaxies inside compact groups. This approach is complementary to group-wide {\Hi} observations \citep[e.g][]{Verdes-Montenegro01} in that we focus on a different phase of the interstellar medium (ISM), and zoom in on the stellar and ionized gaseous content of the galaxies themselves. By targeting galaxies at z$\la$0.03 with a spectral resolution R=7000 and a spectral pixel (spaxel) size of 1 arcsec$^2$ (with natural seeing), we gain access to the physics of the ionized gas on scales of $\sim$1~kpc and simultaneously resolve velocity dispersions down to $\sigma\ga20$ km s$^{-1}$: an ideal combination to detect and study localized consequences of the compact group environment on galaxies. 

Our targets are drawn from both the Hickson and Southern Compact Groups. The Hickson Compact Groups \citep[HCGs,][]{Hickson82a,Hickson82b} represent a class of compact groups that was first identified on Palomar Sky Survey red prints as groups of four or more galaxies with specific brightness, compactness and isolation selection criteria. Most were later on confirmed to be gravitationally bound \citep{Hickson92,Ponman96}. Along with their southern sky cousins \citep[Southern Compact Group or SCG,][]{Iovino02}, the proximity of these low redshift systems \citep[median $z\approx0.03$,][]{Hickson92} make them optimum targets for spatially and spectrally detailed observations.

In \cite{Vogt13}, we focused on NGC~838 in HCG~16, confirmed the presence of a large scale galactic wind and characterized it as young, photoionized and asymmetric, very much unlike the near-by shock-excited galactic wind of NGC~839 \citep[][]{Rich10}. In this article, we turn our attention to HCG~91c, a star forming spiral of which the basic characteristics are given in Table~\ref{table:HCG91c}. In an effort to tie the large scale structure of the group to the local environment within the optical disk of HCG~91c, we supplement our optical IFS observations with {\Hi} observations from the \emph{Very Large Array} (\emph{VLA}) and with broadband optical images from the Pan-STARRS survey. Effectively, we build a multi-phase, multi-scale view of HCG~91c to understand the presence of abundance and velocity offsets in some of the {\Hii} regions detected with WiFeS inside this galaxy.

\begin{table}
\caption{Basic characteristics of HCG~91c}\label{table:HCG91c}
\flushleft{
\begin{tabular}{l p{2.9cm} c}
\hline
\hline
Property & Value & Ref.\\
\hline
Names & HCG~91c &\\
& ESO 467 - G 013 & \\[0.5ex]
R.A. [J2000] & 22$^{h}$09$^{m}$07.7$^{s}$ &\\[0.5ex]
Dec. [J2000] & -27$^{\circ}$48$^{\prime}$34$^{\prime\prime}$  & \\[0.5ex]
Redshift & 0.024414 & (a)\\
                & 0.024377 & (b)\\[0.5ex]
Radial velocity & 7319 km s$^{-1}$ & (a)\\
& 7308 km s$^{-1}$ & (b)\\[0.5ex]
Distance & 104 Mpc &\\[0.5ex]
Spatial scale & 504 pc arcsec$^{-1}$ &\\[0.5ex]
Diameter & 50 arcsec $\approx$ 25 kpc & (b)\\[0.5ex]
R$_{25}$ & 26.75$\pm$3.25 arcsec & (c)\\[0.5ex]
Absolute rotation velocity & 100$\pm$11 km s$^{-1}$ & (b)\\
\quad(at radii$>$22~kpc) & \\[0.5ex]
Star formation rate  & 2.19 M$_{\odot}$ yr$^{-1}$ & (d)\\
                                    & 2.10$\pm$0.06 M$_{\odot}$ yr$^{-1}$ & (b)\\[0.5ex]
Stellar mass & 1.86$\times$10$^{10}$ M$_{\odot}$ & (d) \\[0.5ex]
{\Hi} mass  & 2.3$\times10^{10}$ M$_{\odot}$ & (e)\\
\hline
\end{tabular}
}
(a) \cite{Hickson92}, (b) this work, (c) \cite{Vaucouleurs91}, (d) \cite{Bitsakis14}, (e) \cite{Borthakur10}.
\end{table} 

The article is structured as follows. We first discuss our different datasets (including their acquisition, reduction and processing) in Section~\ref{sec:data}. We then describe the group-wide structure and kinematic of the {\Hi} gas in HCG~91 in Section~\ref{sec:group}, and zoom-in on HCG~91c in Section~\ref{sec:galaxy}. We discuss possible connections between group-wide mechanisms and specific star forming regions inside HCG~91c in Section~\ref{sec:discussion}, and summarise our conclusions in Section~\ref{sec:summary}.

Throughout this paper and unless noted otherwise, when we refer to specific emission lines we mean [N\,{\sc ii}]$\equiv$[N\,{\sc ii}$]\lambda$6583, [S\,{\sc ii}]$\equiv$[S\,{\sc ii}$]\lambda$6717+$\lambda$6731, [O\,{\sc ii}]$\equiv$[O\,{\sc ii}$]\lambda$3727+$\lambda$3729 and [O\,{\sc iii}]$\equiv$[O\,{\sc iii}$]\lambda$5007. We adopt the maximum likelihood cosmology from \cite{Komatsu11}: $H_0$=70.4 km s$^{-1}$ Mpc$^{-1}$, $\Omega_\Lambda$=0.73 and $\Omega_{\rmn{M}}$=0.27, resulting in a distance to HCG~91c of 104~Mpc.

\section{Observational datasets}\label{sec:data}

\subsection{WiFeS}\label{sec:wifes}

\subsubsection{Observations}\label{sec:wifes_obs}
\begin{figure}
\centerline{\includegraphics[scale=0.34]{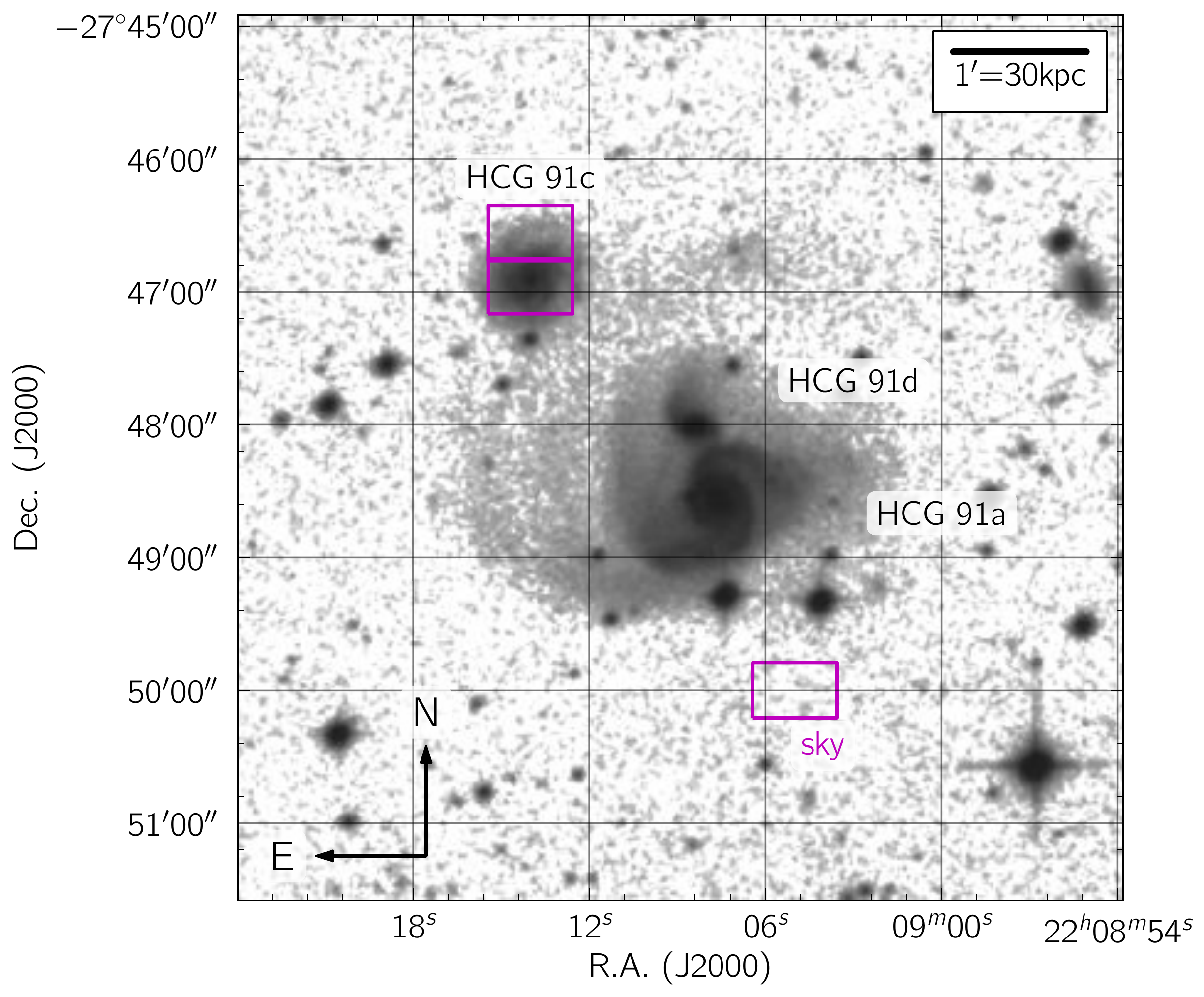}}
\caption{Red band image of HCG~91a, c and d from DSS-2. The purple rectangles denote the footprint of our WiFeS observations of HCG~91c. The location of the sky field was chosen to avoid any of the faint and extended tidal structures stemming from HCG~91a, including the faint tidal arm extending (on-sky) from HCG~91a to HCG~91c.}\label{fig:wifes_chart}
\end{figure}

We observed HCG~91c with WiFeS, the Wide-Field Integral Field Spectrograph \citep{Dopita07,Dopita10} mounted on the 2.3m telescope \citep[][]{Mathewson13} of the Australian National University at Siding Spring Observatory in Northern New South Wales, Australia. The footprint of our observation is shown in Figure~\ref{fig:wifes_chart}, overlaid on a red-band image from the \emph{Second Digitized Sky Survey} (DSS-2)\footnote{ The DSS-2 data was obtained from the European Southern Observatory Online Digitized Sky Survey Server.}. We combined two individual WiFeS pointings to construct a 38$\times$50 arcsec$^2$ mosaic in order to better cover the spatial extent of HCG~91c. 

The data was acquired over two nights on 2012 August 15 and 16, with a seeing of 1.2-1.5 arcsec. Each individual science field of 25$\times$38 arcsec$^{2}$ was observed 4$\times$1400s, resulting in 5600s on-source per field. A blank patch of sky to the South-West of HCG~91a was observed for a total integration time of 700s between every pair of consecutive science frames. 

WiFeS is a dual-beam integral field spectrograph with independent red and blue channels. We used the B3000 grating for the blue arm and the R7000 grating for the red arm, in conjunction with the RT560 dichroic. This setup allows us to have a complete wavelength coverage from 3800{\AA} to 7000{\AA}, as well as a high spectral resolution observation of the H$\alpha$ emission line to perform a detailed kinematic analysis. It is important to note here that with this instrumental setting, the final spectra associated with each spectral pixel (a.k.a spaxel) in the mosaic has a dual spectral resolution of R=3000 from 3800{\AA} to 5560{\AA} and R=7000 from 5560{\AA} to 7000{\AA}, corresponding to a full width at half maximum (FWHM) for emission lines of $\sim$100 km s$^{-1}$ and $\sim$43 km s$^{-1}$, respectively. 

\subsubsection{Data reduction}\label{sec:wifes_reduc}
Each science exposure was individually bias subtracted, flat fielded, sky subtracted, wavelength and flux calibrated, as well as atmospheric refraction corrected with \textsc{pywifes} (v0.5.6), the new official \textsc{python} data reduction pipeline for WiFeS observations \citep{Childress14b,Childress14a}. This new pipeline entirely replaces the now obsolete \textsc{iraf}\footnote{\textsc{iraf} is distributed by the National Optical Astronomy Observatories, which are operated by the Association of Universities for Research in Astronomy, Inc., under cooperative agreement with the National Science Foundation. See http://iraf.noao.edu} pipeline \citep{Dopita10} by providing several key improvements: scriptability, reproducibility, advanced wavelength calibration technique using an accurate model of the instrument's optics, multicore processing and compatibility with the first \& second generations of CCD detectors in the WiFeS cameras. The pipeline is described in detail in \cite{Childress14a}, to which we refer the reader for more details about the different functions. 

The reduced eight red science frames and eight blue science frames are median combined in two mosaics (red and blue) using a custom made \textsc{python} script. Because the wavelength sampling was chosen to be the same during the data reduction process, mosaicking only requires a shift of the data cubes in the spatial x- and y- directions. Given the seeing conditions and the WiFeS spaxel size of 1$\times$1 arcsec$^2$, we restrict ourselves to integer spatial shifts. The final mosaic contains 38$\times$50=1900 spaxels. Once the red and blue mosaics are constructed, we correct them for Galactic extinction using the \cite{Schlafly11} recalibration of the \cite{Schlegel98} extinction map based on dust emission measured by COBE/DIRBE and IRAS/ISSA. We assume for HCG~91c a Galactic extinction A$_V$=0.052, following the NASA/IPAC Extragalactic Database (NED). The recalibration assumes a \cite{Fitzpatrick99} reddening law with $R_V$=3.1 and a different source spectrum than that used by \cite{Schlegel98}. 

We note that the far blue end of a WiFeS spectrum below 4000{\AA} (with the B3000 grating) is a known problematic region. The core reason, aside of a drop of the overall instrument transmission, is that the available flat-field lamps (at the epoch of our observations) have almost no flux at these wavelengths. It is therefore virtually impossible to accurately flat-field the data below 4000{\AA} with the B3000 grating. \textsc{pywifes} combines lamp flat-field images with twilight sky flat-field images which mitigate this issue partially, but not entirely. In the present case, this flat-fielding issue directly affects the [{\Oii}] flux, to which we find a correction factor of 1.5 ought to be applied. This correction factor is only an average for the entire field of view, so that we decided not to rely on the [{\Oii}] line flux in our spaxel-based analysis of HCG~91c. 

\subsubsection{Spectral fitting}\label{sec:wifes_fitting}

We fit the strong emission lines for all 1900 spaxels in our mosaic after fitting and removing the underlying stellar continuum in the spectra. This task is performed automatically using the custom build \textsc{lzifu} v0.3.1 \textsc{idl} routine, written by I-Ting Ho at the University of Hawaii \citep[][Ho et al., in prep.]{Ho14}. \textsc{lzifu} is a modified version of the \textsc{uhspecfit} \textsc{idl} routine to which D.S.N. Rupke, J.A. Rich and H.J. Zahid all contributed. The code is a wrapper around \textsc{ppxf} \citep[][]{Cappellari04}, used to perform the continuum fitting, and \textsc{mpfit} \citep[][]{Markwardt09}, used to fit the emission lines with different gaussian components. 

The \textsc{lzifu} routine first performs (via \textsc{ppxf}) the continuum fitting by looking at spectral regions free of emission lines. We refer the reader to the \textsc{ppxf} documentation for further details. We use the \cite{Gonzales05} set of stellar templates with the Geneva isochrones to construct the best-matched stellar continuum for each spaxel. We note that we tested using the stellar templates based on the Padova isochrones instead, but found no significant difference in the fit results. 

\textsc{lzifu} offers the possibility to fit one or multiple gaussian components to multiple emission lines simultaneously. In the present case, a visual inspection of the data revealed a very narrow structure of the different emission lines with no evidence of multiple peaks in any spaxel. For consistency, we performed two distinct fits of the emission lines - with one, and subsequently with two gaussian components. We detected the presence of a broadened base around the H$\alpha$ emission line in the $\sim$100 brightest spaxels, and confirmed the detection with a statistical f-test \cite[e.g.][]{Westmoquette07}. We performed this test blindly for the [{\Nii}] and H$\alpha$ emission lines only for all spaxels, and with a null-hypothesis rejection probability of 0.01. However, because a) the broadened base is visually only marginally detected around the H$\alpha$ line in $\sim10$ out of 1081 spaxels with signal-to-noise (S/N) larger than 5, b) it is not detected around any other emission line, and c) beam-smearing effects may be responsible for some of the broad component, we choose to enforce a 1-gaussian component fit for every spaxel in our mosaic of HCG~91c. 

We note here that nowhere in our observations do we detect clear multiple peaks in the emission line profile of H$\alpha$ as reported by \cite{Amram03} in their Fabry-Perot observations of HCG~91c. Although the spectral sampling (0.35{\AA}) and spectral resolution (R=9375 at H$\alpha$) of their data are slightly better than ours (0.44{\AA} and R=7000), WiFeS (of which the instrumental design minimizes scattered light and other instrumental artefacts) should have clearly detected the distinct velocity components offset by $\sim$100 km s$^{-1}$ observed by \cite{Amram03}. A detailed comparison of the WiFeS line profile with the velocity structures of \cite{Amram03} is presented in Appendix~\ref{app:amram}. With no evidence for multiple components and/or significant asymmetries in the line profiles in our WiFeS observations,  we are led to conclude that the gas in HCG~91c is associated to a single kinematic structure at all locations.

\begin{figure}
\centerline{\includegraphics[scale=0.4]{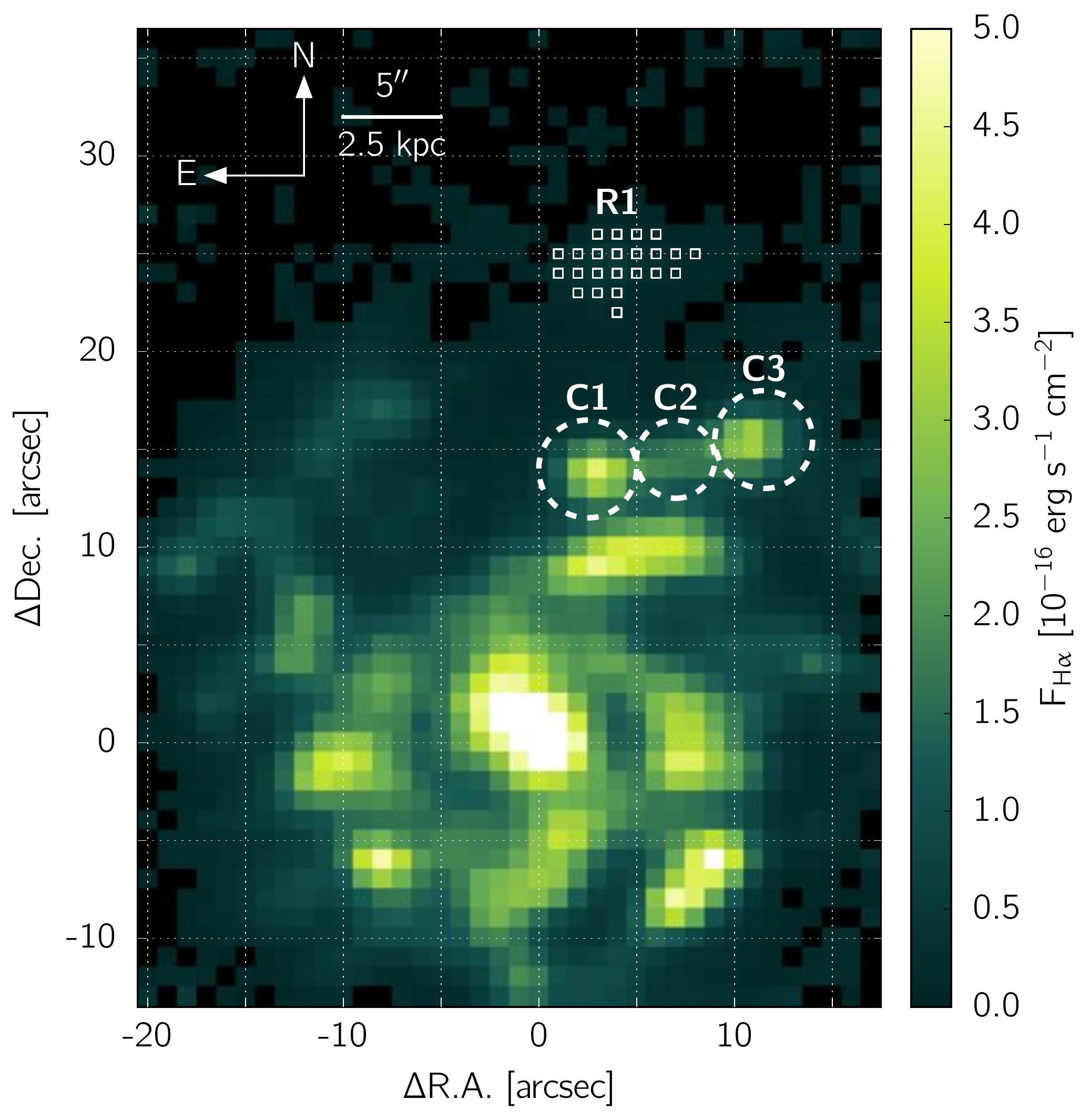}}
\caption{H$\alpha$ flux map of HCG~91c, before correcting the extragalactic reddening. The core of the galaxy and the spiral structure is traced by enhanced H$\alpha$ emission. Three star forming regions of interest ``C1'', ``C2'' and ``C3'' are labelled accordingly. The region ``R1'' is also indicated, along with the individual spaxels it comprises. The x and y axis are in arcsec ($\equiv$ spaxels), centered on the brightest spaxel in H$\alpha$. At the distance of HCG~91c, 1 arcsec $\approx$ 503 pc.}\label{fig:ha_map}
\end{figure}

The H$\alpha$ flux map of HCG~91c observed by WiFeS is shown in Figure~\ref{fig:ha_map} in units of $10^{-16}$ erg s$^{-1}$ cm$^{-2}$. The final emission line widths quoted in this article are corrected for the instrumental resolution. 

Throughout this article, we analyze our WiFeS dataset on a \emph{per-spaxel} basis to keep a high spatial resolution even at large distances from the galaxy center. To ensure trustworthy results, we adopt varying S/N cuts throughout our analysis. For example, we require S/N(H$\alpha$;H$\beta$)$\geq$5 to perform the extragalactic reddening correction. The specific S/N cuts will be discussed individually for each case in the next Sections. For clarity, we also include the associated S/N cut and number of spaxel in all relevant diagrams throughout this article.

A series of spaxels with S/N(H$\alpha$)$\geq$5 and S/N(H$\beta$)$<$5 are detected to the North of HCG~91c at the mean position [4;25]. This is the most distant star forming region (from the galaxy center) detected with WiFeS, and as such it holds unique clues regarding the state of the ionized gas at large radii. To ensure a reliable extragalactic correction, we have summed all of the contiguous spaxels with S/N(H$\alpha$)$\geq$5 into a combined spectrum, which we refer to as the ``R1'' region.  The individual R1 spaxels are marked with empty white squares in Figure~\ref{fig:ha_map}. We also indicate the location of three star forming regions with anomalous abundance and kinematics offsets, the ``C1'', ``C2'' and ``C3'' regions. 

\subsubsection{Extragalactic reddening}
\begin{figure}
\centerline{\includegraphics[scale=0.4]{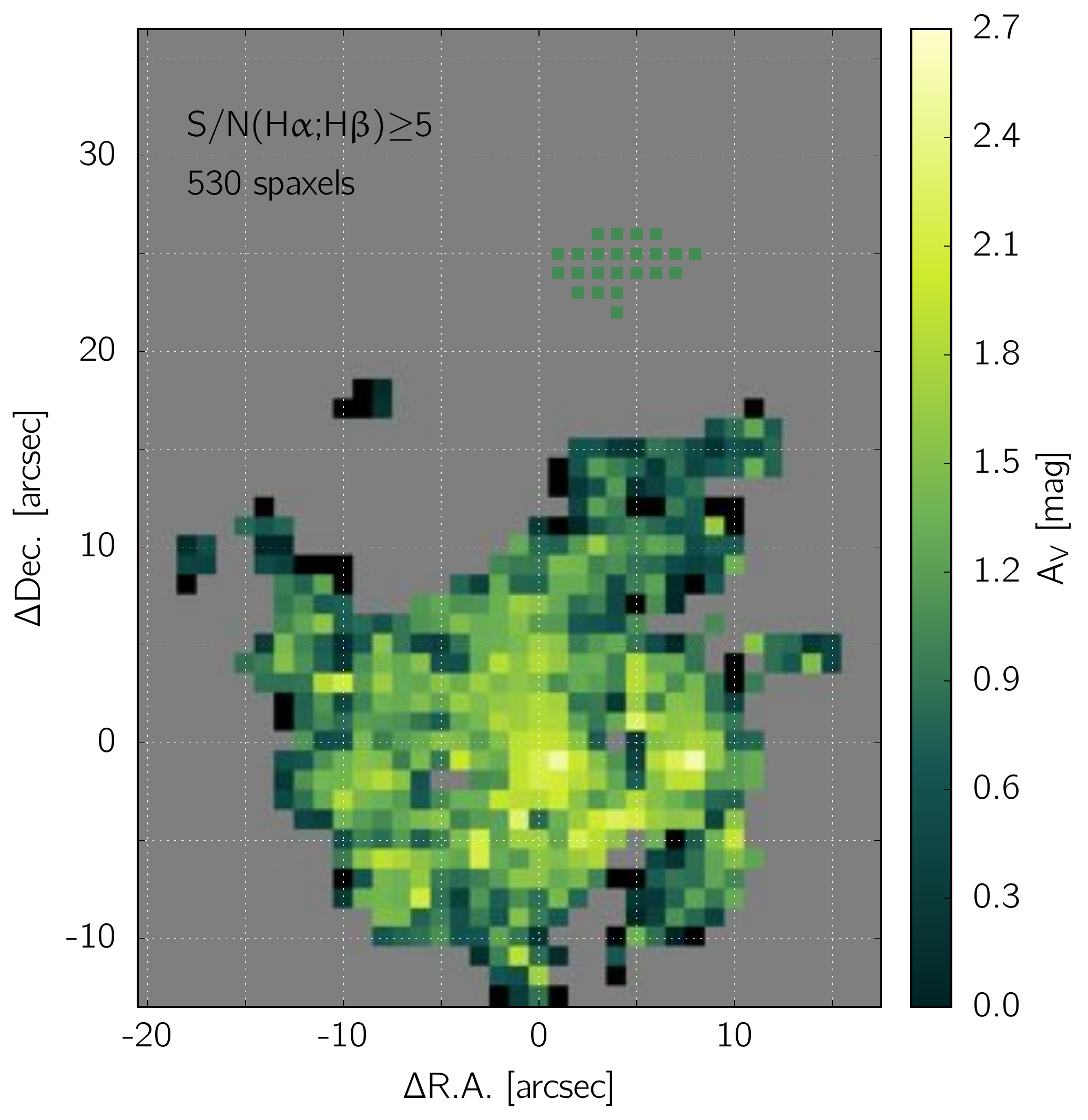}}
\caption{A$_\text{V}$ extinction, calculated from the H$\alpha$/H$\beta$ emission line flux ratio. The center of the galaxy is the most affected by extragalactic reddening, with A$_\text{V}>2$. Enhanced extinction regions are also associated with the spiral structure detected in Figure~\ref{fig:ha_map}. The uncertainty in the A$_\text{V}$ value is of the order of 0.6-0.8 for all spaxels. }\label{fig:av}
\end{figure}

We correct our WiFeS observations for extragalactic reddening using the H$\alpha$/H$\beta$ line ratio. HCG~91c does not contain any active galactic nuclei (AGN) which could increase the intrinsic line ratio value $R_{\alpha\beta}$, and we therefore assume that $R_{\alpha\beta}$ = 2.86 as in a case B recombination \citep[$n_e=100$ cm$^{-1}$; T=$10^{4}$ K, see][]{Osterbrock89}. We follow the methodology described in details in the Appendix A of \cite{Vogt13}, and use the extinction curve from \cite{Fischera05} with R$_V^A$=4.5. The choice of this extinction curve, very similar to that of \cite{Calzetti00}, is motivated by the fact that \cite{Wijesinghe11} showed that for galaxies in the GAMA survey \citep{Driver09}, it provides self-consistent estimates of the star formation rates computed from the UV, H$\alpha$ and [{\Oiii}]. The resulting V-band extinction A$_\text{V}$ in magnitudes is shown in Figure~\ref{fig:av}. The extragalactic reddening (and correction) is only calculated (and applied) for the 530 spaxels with S/N(H$\alpha$;H$\beta$)$\geq$5, corresponding to a line intensity detection threshold of $\sim$2$\times$10$^{-17}$ erg s$^{-1}$ cm$^{-2}$ {\AA}$^{-1}$.

The largest extinction is present towards the core of the galaxy with A$_\text{V}>2$. Further out, the extinction drops to A$_\text{V}$$\approx$1, with localized increases coherent with the spiral arms visible in Figure~\ref{fig:ha_map}. Unlike most star forming regions in the spiral arms, we detect no enhancement of the extinction associated with the C1, C2 and C3 regions at the 1-sigma level (given an uncertainty of 0.6-0.8 in the A$_\text{V}$ value).

\subsection{\emph{VLA}}\label{sec:vla}

The {\Hi} distribution in HCG~91 was observed with the \emph{Very Large Array} (\emph{VLA}) under the program AV0285 (P.I.: L. Verdes-Montenegro) on 2005 October 5. The observations were carried-out in the \emph{VLA} D-configuration with 3.25hr of on-source integration and dual polarization using two intermediate frequency (IF). The resulting total bandwidth of the datacube is 6.2~MHz ($\sim$1200 km s$^{-1}$) with a resolution of 97~kHz. The correlator was setup to cover the entire {\Hi} spectral linewidth with a velocity resolution of $\sim$21.6 km s$^{-1}$. The data was calibrated following the standard \emph{VLA} calibration and imaging procedures using NRAO's Astronomical Image Processing Software (AIPS). The final absolute uncertainty in the flux density scaling is $\sim$10 per cent. The synthesized beam of the reduced datacube is 74"$\times$47", with a position angle of 4.83$^{\circ}$ West-of-North \citep{Borthakur10}.

\subsection{Pan-STARRS}\label{sec:panstarrs}

HCG~91c was observed at multiple epochs with the Pan-STARRS\footnote{ Panoramic Survey Telescope And Rapid Response System}  1 (PS1) telescope \citep{Kaiser02, Kaiser04,Kaiser10, Hodapp04} on Haleakala (Maui) as part of the PS1 3$\pi$ survey (Chambers et al., in preparation). This survey mapped (several times) the entire sky visible from Hawaii ($\delta$$>$-30$^{\circ}$) in five broad filters \citep[g$_{\rmn{P1}}$:$\sim$400-550 nm; r$_{\rmn{P1}}$:$\sim$550-700 nm; i$_{\rmn{P1}}$: $\sim$690-820 nm; z$_{\rmn{P1}}$:$\sim$820-920 nm; y$_{\rmn{P1}}$:$\sim$920-1000 nm, see][]{Tonry12} between 2010 and 2014. The Pan-STARRS images presented in Section~\ref{sec:pans} were obtained from the online Postage Stamp server. These images were automatically processed by the PS1 Image processing Pipeline \citep[IPP, see e.g.][]{Magnier07, Magnier13}. Specifically, the Pan-STARRS images presented in this article (see Figure~\ref{fig:panstarrs}) have been obtained from the 3$\pi$ survey processing version 2 (PV2) release. They have a pixel size of 0.25 arcsec, and are the result of the combination of 20 single epoch exposures obtained under natural seeing conditions ($\sim$1 arcsec). 

\section{The group-wide {\Hi} distribution in HCG~91}\label{sec:group}

\begin{figure*}
\centerline{\includegraphics[scale=0.23, ]{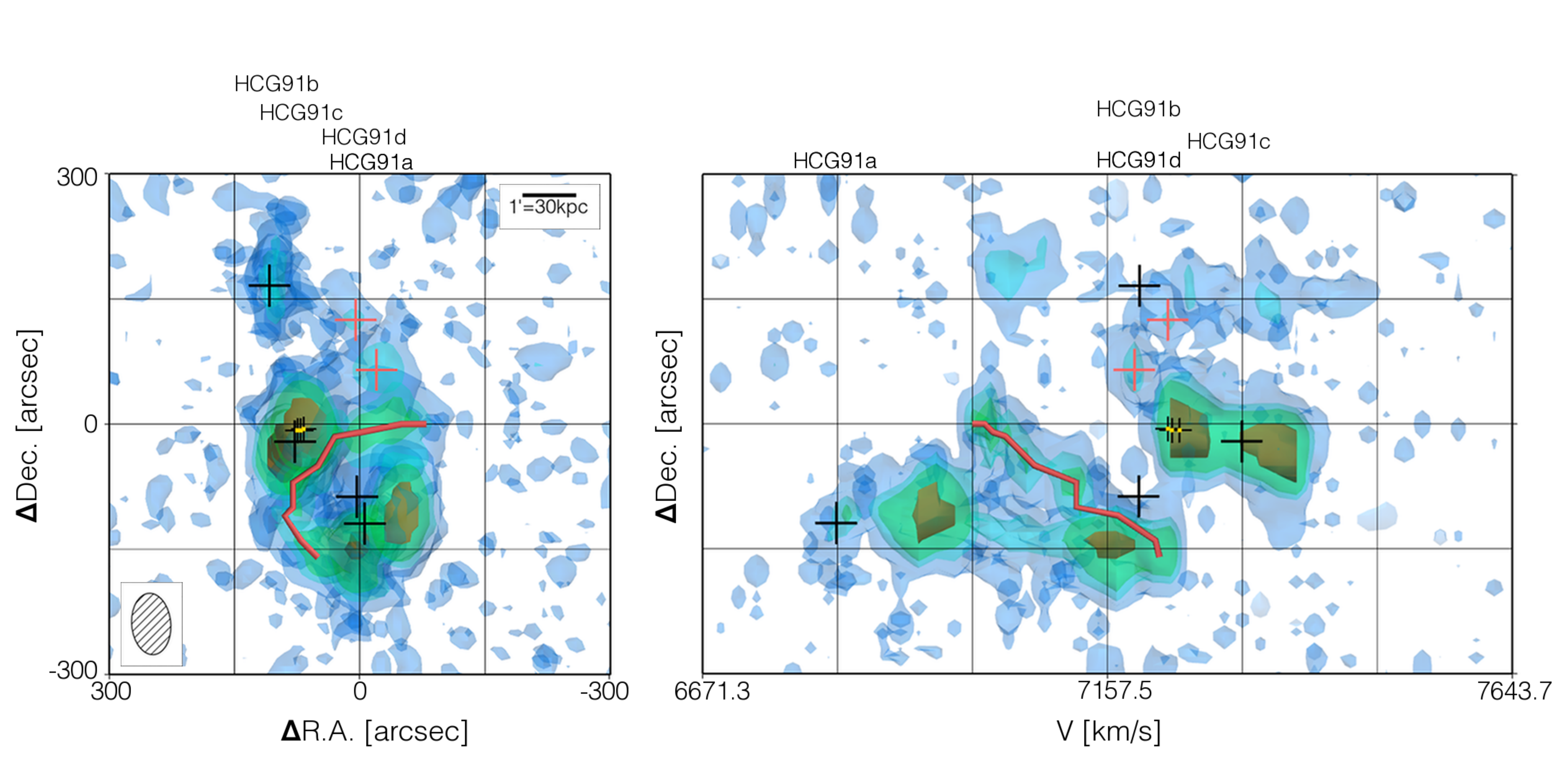}}
\caption{Position-Position (left) and Position-Velocity (right) projection of the {\Hi} gas content of HCG~91, as seen by the \emph{VLA}. The gas density is traced by a series of iso-intensity semi-transparent surfaces (set at 1.3, 2.5, 3.5 and 6.0 mJy/beam, respectively). The location of the different group members is marked with black crosses, and labelled accordingly above either panels. Two {\Hi} clumps associated with the main gas reserve around HCG~91c are marked with red crosses for unambiguous identification.  The large tidal tail originating to the South-East of HCG~91a is traced by a red rod. The C1, C2 and C3 star forming regions in HCG~91c are marked with golden cubes and black crosses (the golden cubes are 5 arcsec $\times$ 5 arcsec $\times$ 5 km s$^{-1}$ in size). The corresponding \emph{VLA} beam size \citep[74''$\times$47'', see][]{Borthakur10} is shown to the bottom left of the left panel. \st{An interactive version of this Figure can be accessed either with Adobe Acrobat Reader v.9.0 or above by clicking on this Figure, or via most mainstream web browsers by opening the corresponding \textsf{HTML} file included as supplementary material to this article. The 3-D cube can be freely rotated and/or zoomed in and out. In the interactive cube, the blue axis is the velocity dimension, the red axis is the R.A. direction, and the green axis the Dec. direction. A series of \emph{pre-set} views allow the viewer to easily reproduce projections of interest or remove the lowest intensity iso-surface for a clearer view of each galaxy. The golden cubes marking the location and spatial/kinematic extent of the C1, C2 and C3 star forming regions are best seen in the interactive counterpart of this Figure.} \textit{NdlA: the interactive component of this Figure has been disabled in the arXiv version of the article (its compilation requires the media9 package, while arXiv currently only supports the obsolete movie15 package). The interactive Figure will be accessible from the final article in MNRAS (both as an interactive \textsf{PDF} inside the article itself, and as a supplementary interactive \textsf{HTML} document). Until then, readers can access the interactive \textsf{HTML} version of this Figure at \url{http://www.mso.anu.edu.au/~fvogt/website/misc.html}.} }\label{fig:HI_3d}
\end{figure*}

{\Hi} is an excellent tracer of past and ongoing gravitational interactions. For compact groups, {\Hi} can also be used to estimate the evolutionary stage of the group overall \citep{Verdes-Montenegro01}. The {\Hi} map for HCG~91 obtained with the \emph{VLA} is shown in Figure~\ref{fig:HI_3d}. The left and right panels show Position-Position and Position-Velocity projections of the data cube, respectively. Iso-intensity surfaces are fitted to the distribution of {\Hi} emission in the 3-dimensional datacube using the \textsc{mayavi} module in \textsc{python} \citep{Ram11}, before being projected onto the respective 2-dimensional planes. Each iso-intensity level is shown with an 80 per cent transparency, with exception to the inner-most one with 0 per cent transparency. The locations of the HCG~91a, b, c and d galaxies are marked with black crosses. The location of the C1, C2 and C3 star forming regions in HCG~91c is indicated with small golden cubes and black crosses. {\Hi} structures of interest (see below) are marked with red-crosses for rapid identification. A large {\Hi} tail associated with HCG~91a is traced with a red rod. Figure~\ref{fig:HI_3d} is also interactive, following a concept described by \cite{Barnes08}. By clicking on it in Adobe Acrobat Reader v9.0 or above, it is possible to load an interactive view of the (X;Y;v) data cube. The interactive model is also accessible via a supplementary \textsf{HTML} file compatible with most mainstream web browsers \footnote{At the date of publication, the interactive \textsf{HTML}document is compatible with \textsc{firefox}, \textsc{chrome}, \textsc{safari} and \textsc{internet explorer}. We refer the reader to the X3DOM documentation for an up-to-date compatibility list: http://www.x3dom.org/}. When inspecting this interactive 3-D map, one should remember that it is not ``fully'' spatial, but rather an X-Y-v volume. Presenting the \emph{VLA} 3-D data cube as 3-D model is similar to the example provided by \cite{Kent13}, although our respective 3-D model creation methods are different (i.e. we do not use \textsc{blender}). The total {\Hi} mass associated with HCG~91 is 2.3$\times$10$^{10}$M$_{\odot}$ \citep[][]{Borthakur10}. 

\begin{figure}
\centerline{\includegraphics[scale=0.33]{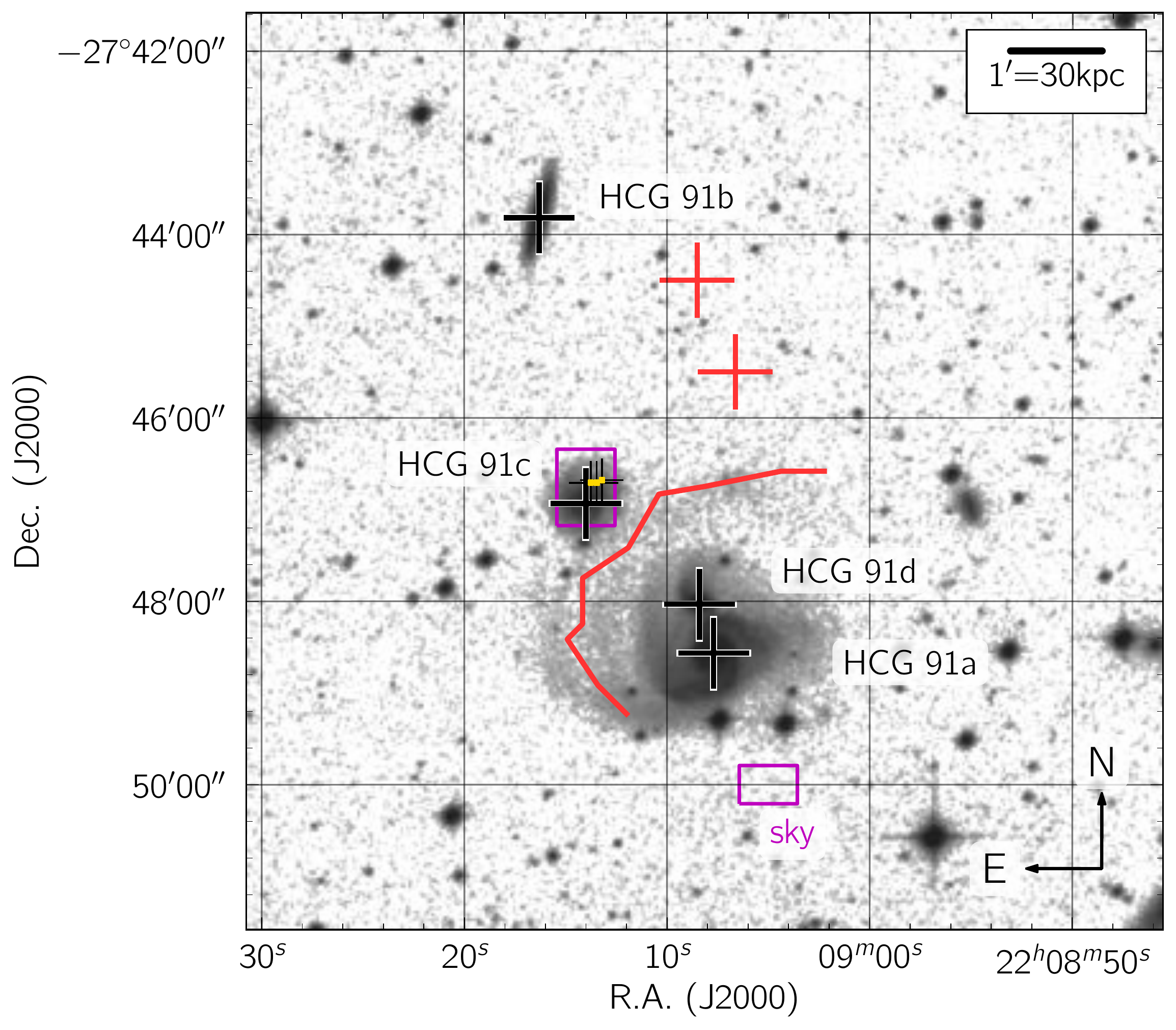}}
\caption{Red band image of HCG~91a,b,c and d from DSS-2. The image spans the same area as the Position-Position velocity diagram of Figure~\ref{fig:HI_3d} (left), of which the same markers are shown for comparison. The purple rectangles denote the footprint of our WiFeS observations of HCG~91c.}\label{fig:dss2-symbols}
\end{figure}

\begin{figure}
\centerline{\includegraphics[scale=0.33]{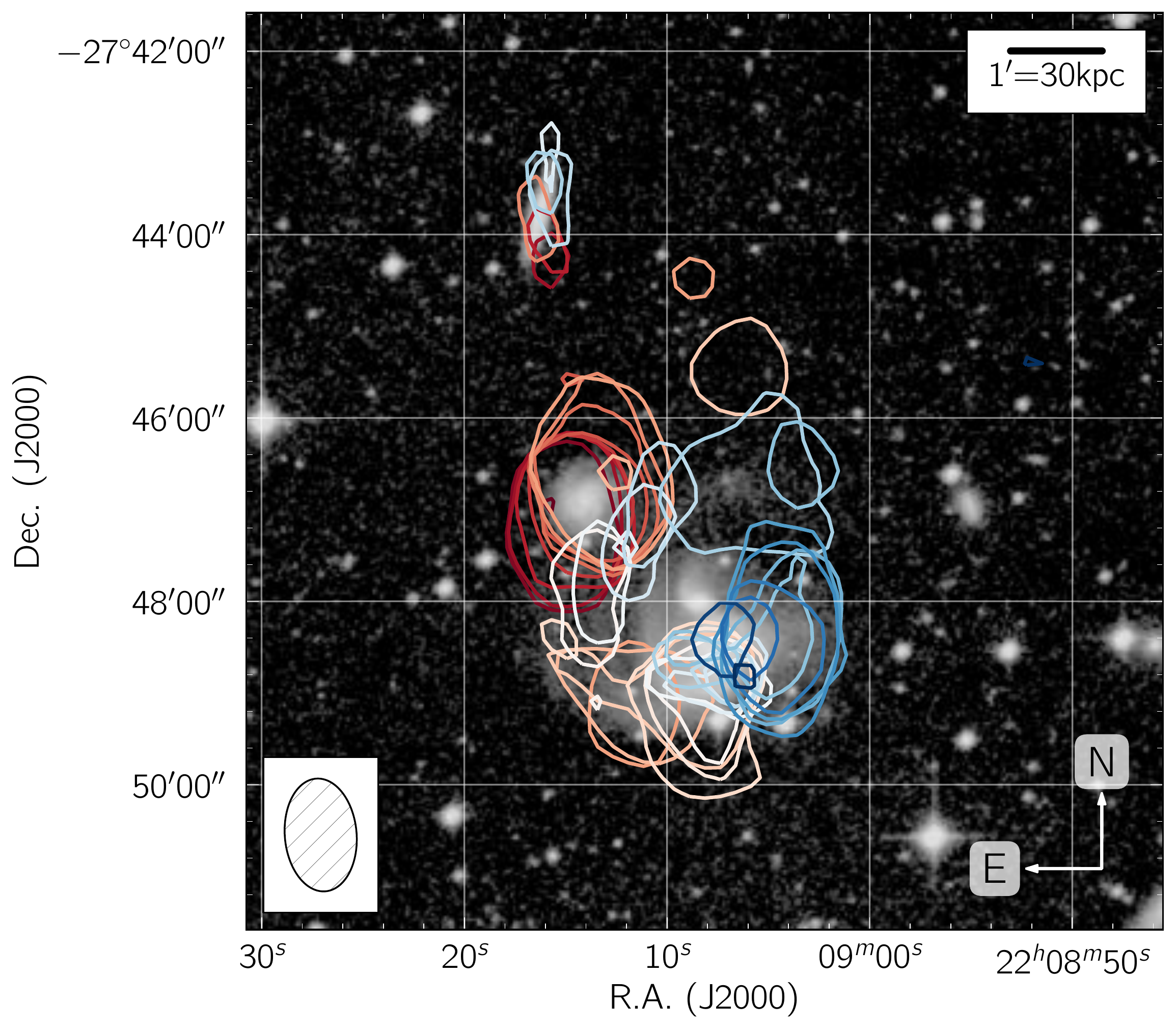}}
\caption{Red band image of HCG~91a,b,c and d from DSS-2. The image spans the same area as shown in Figure~\ref{fig:HI_3d} (left) and \ref{fig:dss2-symbols}. Iso-intensity contours trace the {\Hi} content of the group, and are color-coded as a function of their radial velocity, ranging from 6822.0 km s$^{-1}$ (dark blue) to 7405.5 km s$^{-1}$ (dark red), in steps of $\sim$21.6 km s$^{-1}$. The contours are extracted from the different velocity slices of the \emph{VLA} datacube and trace an intensity of 2.5 mJy/beam. }\label{fig:dss+VLA}
\end{figure}

For ease of visualisation, we show in Figure~\ref{fig:dss2-symbols} the different symbols marking the position of the galaxies, {\Hi} clumps of interest and tidal arm introduced in Figure~\ref{fig:HI_3d} on top of a DSS-2 red band image, along with our WiFeS observation fields. The full extent of the different {\Hi} structures is traced with iso-contours (at a level of 2.5 mJy/beam and colored as a function of the gas velocity) in Figure~\ref{fig:dss+VLA}. 

Globally, the different sub-structures of the {\Hi} gas distribution in the group can be associated with the different individual galaxies. Specifically: 
\begin{itemize}

\item {HCG~91a:} Spatially, this galaxy is coherent with the large {\Hi} structure to the South of the group. The large tidal feature detected in {\Hi} is also spatially coherent with a faint optical counterpart distinguishable in the red band DSS-2 image of the area (see Figure~\ref{fig:dss2-symbols}). Kinematically however, we find a mismatch of $\sim$220 km s$^{-1}$ between the optical redshift of the galaxy \citep[6832 km s$^{-1}$, see][]{Hickson92} and the mean redshift of the {\Hi} structure. Although the optical redshift of HCG~91a associates the galaxy with a local {\Hi} over-density in the Position-Velocity diagram of Figure~\ref{fig:HI_3d}, most of the {\Hi} gas spatially-coherent with the galaxy (including the base of the tidal tail) is redshifted by up to $\sim$400 km s$^{-1}$ with respect to the optical redshift. The {\Hi} distribution observed with the \emph{VLA} (both the spatial extent and the dynamic range of the different structures) is consistent with similar observations from \cite{Barnes01} using the \emph{Australia Telescope Compact Array} (\emph{ATCA}), while the optical redshift measurement of HCG~91a from \cite{Hickson92} (z=0.022789) is consistent with the redshift measurement from the 6dF Galaxy Survey data release 3 \citep[z=0.022739, see][]{Jones04,Jones09} and with our own observations of the galaxy with WiFeS. Altogether, these observations confirm that the observed optical-radio redshift mismatch of HCG~91a is real. Here, we merely mention the existence of this redshift mismatch, and defer any further discussion to a later paper dedicated to our WiFeS observations of HCG~91a.

\item{HCG~91b:} A spatially compact and kinematically extended ($\sim$490 km s$^{-1}$) {\Hi} structure is associated with this galaxy. The large range of the {\Hi} kinematics is consistent with the fact that HCG~91b is seen almost edge-on on.

\item{HCG~91c:} A rotating {\Hi} disk with a velocity range of $\sim$200 km s$^{-1}$ ($\sim$7200-7400 km s$^{-1}$) is associated both spatially and kinematically with the optical counterpart of HCG~91c. We find a small kinematic offset of $\sim$10 km s$^{-1}$ between the optical redshift of HCG~91c measured by \cite{Hickson92} and the mean velocity of the {\Hi} structure. The largely undisturbed morphology of the {\Hi} distribution is suggesting the presence of only minimal tidal effects for this galaxy. To the North-West, two fainter {\Hi} sub-structures (marked with red crosses in Figure~\ref{fig:HI_3d}) appear connected to the main gas reservoir of HCG~91c. They are also connected (less strongly) at the 1.3 mJy/beam level to the {\Hi} structure associated with HCG~91b. These two {\Hi} clumps are located 116 arcsec $\approx$ 58.4 kpc and 150 arcsec $\approx$ 75.5 kpc from the center of HCG~91c. They have no visible optical counterpart in the DSS-2 red band image in Figure~\ref{fig:dss2-symbols}. Spatially, HCG~91c is located $\sim$15 kpc to the North-East of the large tidal arm originating from HCG~91a. The {\Hi} gas in the tidal arm is blueshifted by 150-250 km s$^{-1}$ from the mean velocity of HCG~91c.

\item{HCG~91d:} This galaxy is not associated with any {\Hi} structure kinematically.
\end{itemize}

The large tidal feature originating from HCG~91a makes the HCG~91 compact group a \emph{Phase 2} group in the classification of \cite{Verdes-Montenegro01}, although some of the {\Hi} gas is still clearly associated with the galaxies HCG~91b and HCG~91c. The {\Hi} reservoir associated with HCG~91c appears largely undisturbed from a kinematic point of view. The two {\Hi} gas clumps located to the North-West of HCG~91c may have resulted from tidal stripping, suggesting that HCG~91c is experiencing the first stages of tidal disruption via gravitational interaction. The {\Hi} {\it bridge} at the 1.3 mJy/beam level connecting the gas reservoir of HCG~91b and HCG~91c could be seen as evidence for an ongoing interaction between the {\Hi} envelopes of these two galaxies, although the exact {\it bridge} structure would require a higher spatial sampling to be clearly established.

\section{The galaxy HCG~91c}\label{sec:galaxy}
Here, we describe the different characteristics of HCG~91c as seen by WiFeS and Pan-STARRS. We focus our analysis on the strong emission lines and the associated underlying physical characteristics of the ionized gas. We restrict ourselves to a description of the system, and postpone a global discussion of the different measurements until Section~\ref{sec:discussion}.

\subsection{Spatial distribution of the stellar light}\label{sec:pans}

\begin{figure*}
\centerline{\includegraphics[scale=0.42]{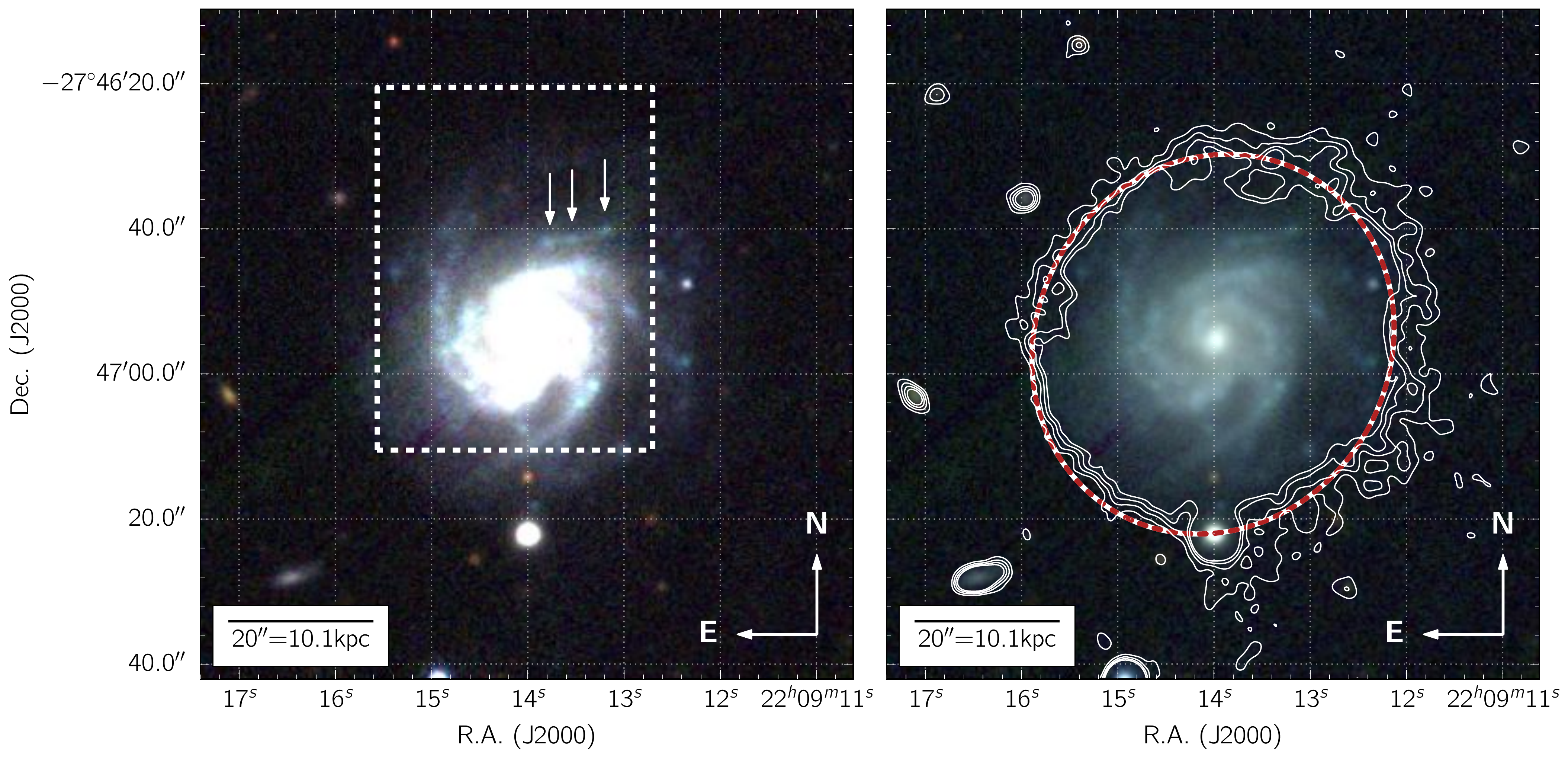}}
\caption{Color combination of Pan-STARRS i$_{\rmn{P1}}$, r$_{\rmn{P1}}$ and g$_{\rmn{P1}}$ images (shown as the red, green and blue channels respectively) of HCG~91c. The color scheme is stretched to reveal the galaxy outskirts (left) or the inner region and spiral structure (right). The WiFeS observations footprint is marked with a dashed rectangle in the left panel, along with three arrows marking the location of the C1, C2 and C3 regions. R$_{25}$ is traced by the dashed ellipse in the right panel. Also in the right panel, low level iso-contours constructed from a gaussian smoothed g$_{\rmn{P1}}$ band image highlight the full optical extent of HCG~91c on the sky at a distance of $\sim$R$_{25}$.}\label{fig:panstarrs}
\end{figure*}

A color-composite image of HCG~91c, constructed from the Pan-STARRS i$_{\rmn{P1}}$, r$_{\rmn{P1}}$ and g$_{\rmn{P1}}$ band images, is shown in Fig.~\ref{fig:panstarrs}. The same image is shown twice, with the intensity stretch adjusted to reveal the outer regions (left panel) and the inner spiral structure (right panel) of HCG~91c. The WiFeS mosaic footprint is traced by a dashed rectangle in the left panel, and the location of R$_{25}$ is traced by an ellipse in the right panel. Specifically, R$_{25}$=26.75$\pm$3.25 arcsec =13.5$\pm$0.2 kpc with an ellipticity of 0.91, following NED and the \emph{Third Reference Catalogue of Bright Galaxies} \citep{Vaucouleurs91}. We note that this value is consistent with the estimate of R$_{25}$=28.45 arcsec and an ellipticity of 0.87 from the \emph{Surface Photometry Catalogue of the ESO-Uppsala Galaxies} \citep{Lauberts89}.

HCG~91c harbors a regular, tightly wrapped spiral pattern with a bright nucleus. The outer limit of the stellar disk, traced by white iso-contours in the right panel of Fig.~\ref{fig:panstarrs} (extracted from the g$_{\rmn{P1}}$ image) is mostly regular, with the brightness dropping the most sharply towards the SE. The spiral arms are extending (at least) 20 arcsec $\approx$ 10~kpc from the galaxy center, but their brightness decreases rapidly beyond $\sim$5~kpc. 

The C1, C2 and C3 regions defined in Figure~\ref{fig:ha_map} appear as three distinct compact features (each indicated by a white arrow in the left panel of Figure~\ref{fig:panstarrs} for clarity). The Pan-STARRS broad-band images primarily trace the stellar content of HCG~91c, and one must use caution when comparing this view to that of the ionized gas provided by WiFeS. Nevertheless, with a diameter of $\sim6$ Pan-STARRS pixels $\approx$ 1.5 arcsec $\approx$ 750~pc across, it is very likely that the optical emission associated with the C1, C2 and C3 star forming regions is unresolved both in our WiFeS observations and in the Pan-STARRS image of HCG~91c.

\subsection{Emission line maps and line ratio diagrams}

\begin{figure*}
\centerline{\includegraphics[scale=0.35]{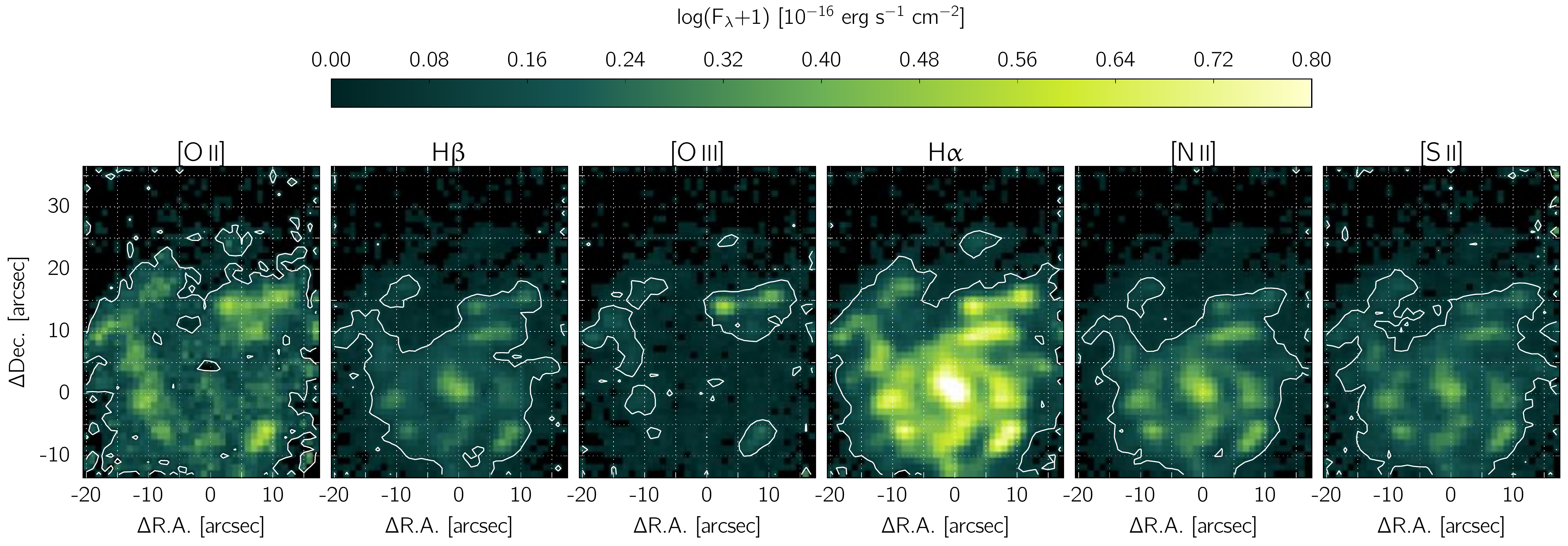}}
\caption{Maps for the different emission lines considered in this article, shown with a uniform color scheme and color stretch for comparison purposes. For each line, the white contours at 2.2$\times$10$^{-17}$ erg s$^{-1}$ cm$^{-2}$ approximatively trace the regions with S/N$\geq$5.}
\label{fig:line_maps}
\end{figure*}

We present the flux line maps for the principle optical emission lines ([{\Oii}], H$\beta$, [{\Oiii}], H$\alpha$, [{\Nii}], [{\Sii}]) in the spectrum of HCG~91c (before applying the extragalactic reddening correction) in Figure~\ref{fig:line_maps}. For comparison purposes, all the maps are shown with the same color stretch and cuts. In each panel, the white contours at 2.2$\times$10$^{-17}$ erg s$^{-1}$ cm$^{-2}$ trace approximatively the regions with S/N$\geq$5. 

The structure of the emission from the [{\Nii}] and [{\Sii}] lines is similar to H$\alpha$, although with an overall weaker intensity. There is a distinct lack of [{\Oii}] emission from the core of the galaxy, a region subject to a larger reddening (see Figure~\ref{fig:av}). The noise level is much higher in the [{\Oii}] map, because the lines are located at the very blue end of the WiFeS spectra (in a region subject to flat-fielding problems, see Section~\ref{sec:wifes_reduc}). The spatial distribution of the [{\Oiii}] line is distinctly different from that of H$\alpha$. The lack of emission at the core of the galaxy is reminiscent of [{\Oii}], and is consistent both with a larger extinction and with a higher gas metallicity overall in the core of HCG~91c (see Section~\ref{sec:pyqz}). Further out, the [{\Oiii}] emission is the strongest in the C1 and C3 regions. By comparison, the H$\alpha$ emission associated with the C1 and C3 regions is at a level comparable to other star forming regions within the spiral arms.

\begin{figure}
\centerline{\includegraphics[scale=0.4]{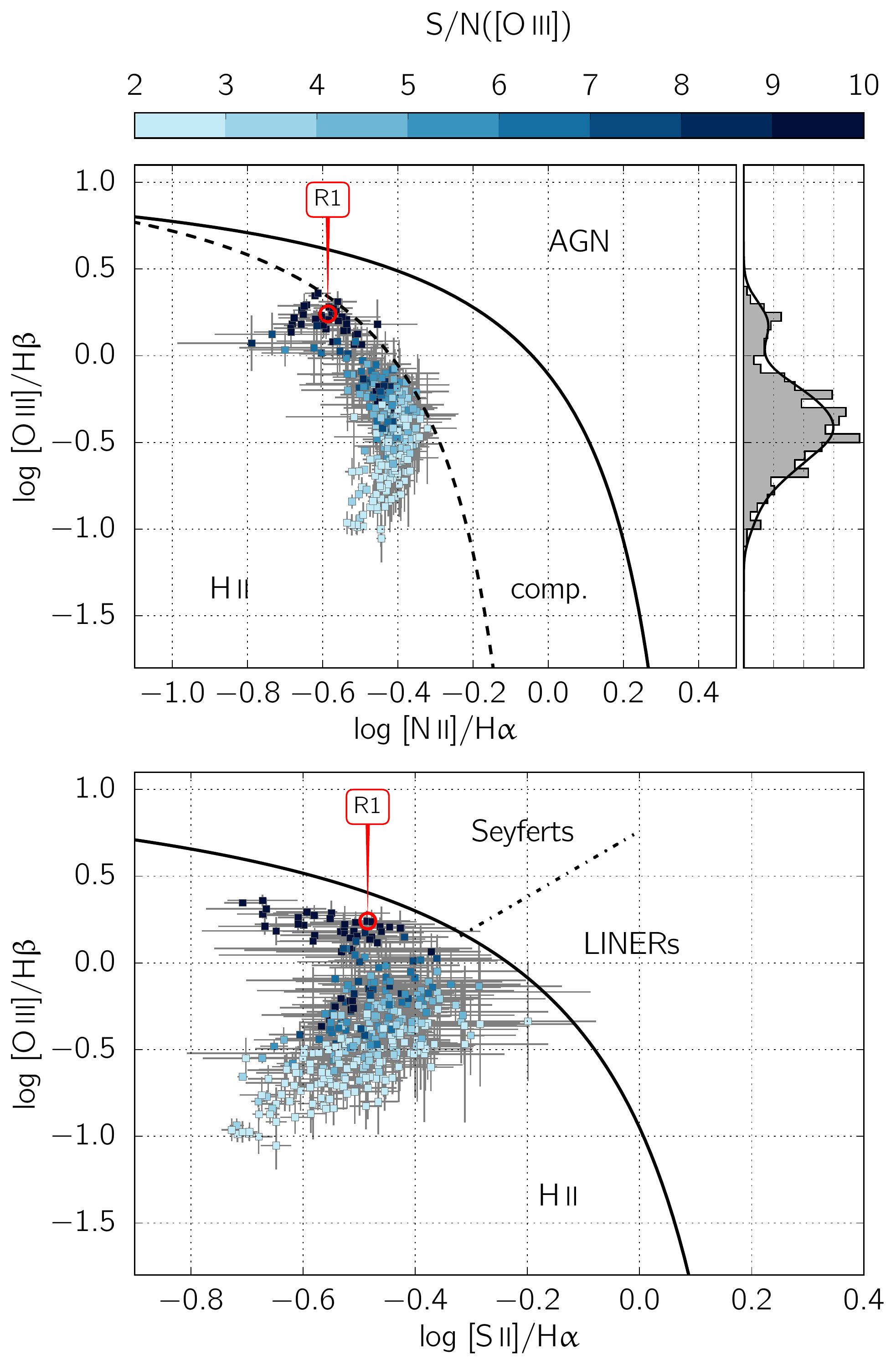}}
\caption{Standard line ratio diagnostic diagrams $\log$~[{\Nii}]/H$\alpha$ \emph{vs.} $\log$~[{\Oiii}]/H$\beta$ (top) and $\log$~[{\Sii}]/H$\alpha$ \emph{vs.} $\log$~[{\Oiii}]/H$\beta$ (bottom). Each spaxel with S/N(H$\alpha$;H$\beta$)$\geq$5 and  S/N([{\Oiii}];[{\Nii}];[{\Sii}])$\geq$2 are marked with squares colored as a function of S/N([{\Oiii}]). The position of the R1 integrated spectra is marked with a red circle and labelled accordingly. The 1-sigma errors are shown in grey for each measurement. The full black lines are the maximum starburst diagnostic lines from \protect\cite{Kewley01}, the dashed line is the starburst diagnostic line from \protect\cite{Kauffmann03}, while the Seyferts-LINERs separation from \protect\cite{Kewley06} is marked as a dash-dotted line in the bottom panel. The histogram in the upper panel shows the spaxel distribution as a function of $\log$~[{\Oiii}]/H$\beta$, while the smooth distribution (traced by the black curve) was derived by performing a Kernel Density Estimation on the distribution of  $\log$~[{\Oiii}]/H$\beta$ values.}
\label{fig:line_ratio}
\end{figure}

The standard line ratio diagrams $\log$~[{\Nii}]/H$\alpha$ \emph{vs} $\log$~[{\Oiii}]/H$\beta$ and $\log$~[{\Sii}]/H$\alpha$ \emph{vs} $\log$~[{\Oiii}]/H$\beta$ for HCG~91c are shown in Figure~\ref{fig:line_ratio}. Each 353 spaxels for which S/N(H$\alpha$;H$\beta$)$\geq$5 and S/N([{\Oiii}];[{\Nii}];[{\Sii}])$\geq$2 are shown as individual squares, color-coded as a function of S/N([{\Oiii}]). The first S/N selection condition results from the extragalactic reddening correction described previously. A lower S/N cut for [{\Oiii}], [{\Nii}] and [{\Sii}] is then acceptable given that both H$\alpha$ and H$\beta$ are clearly detected and strongly constrain the velocity and velocity dispersion of all the different emission lines. The color scheme choice reflects the fact that S/N([{\Oiii}])$\geq2$ is in all cases the \emph{limiting condition}. The grey error bars shown in each diagram indicate the 1-sigma error associated with the line ratio measurements, calculated from the errors measured by \textsc{lzifu} based on the variance measurements propagated through the \textsc{pywifes} data reduction pipeline.   

Within the errors, the different spaxels form a well defined sequence in both line ratio diagrams, extending 1.5dex along the $\log$~[{\Oiii}]/H$\beta$ direction. The spaxel density along the sequence is continuous, except for a marked decrease at $\log$~[{\Oiii}]/H$\beta$$\approx$0. This gap, best seen in the density histogram in the right-hand side of the upper panel of Figure~\ref{fig:line_ratio}, is most certainly not a consequence of low S/N measurements. Indeed, spaxels on either side and across this gap in the distribution have strong oxygen emission with S/N([{\Oiii}]$\ga$6, and are therefore clearly detected.

\begin{figure}
\centerline{\includegraphics[scale=0.4]{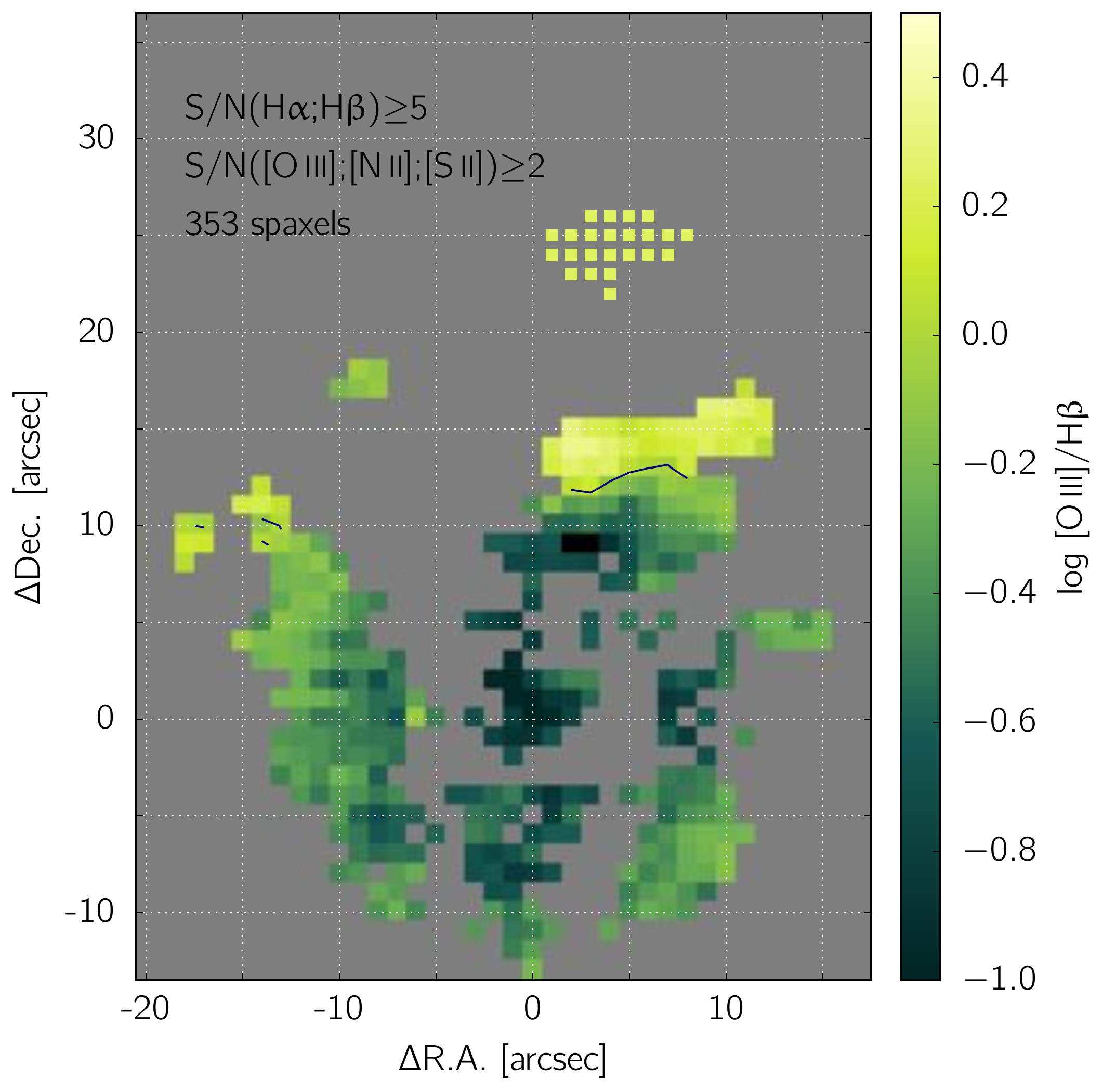}}
\caption{Map of $\log$~[{\Oiii}]/H$\beta$ for spaxels with S/N(H$\alpha$;H$\beta$)$\geq$5 and S/N([{\Oiii}];[{\Nii}];[{\Sii}])$\geq$2. Most spaxels with a positive ratio are located in the C1, C2 and C3 regions. The black line traces the iso-contour corresponding to $\log$~[{\Oiii}]/H$\beta$=0.}\label{fig:o3_hb}
\end{figure}

In Figure~\ref{fig:o3_hb}, we show the $\log$~[{\Oiii}]/H$\beta$ line ratio map for all 353 spaxels visible in either panel of Figure~\ref{fig:line_ratio}. Along with the R1 region, the spaxels associated with the C1, C2 and C3 regions have $\log$~[{\Oiii}]/H$\beta$$\geq$0. They are forming the upper end of the star forming sequence visible in the line ratio diagrams. Altogether, Figures~\ref{fig:line_ratio} and \ref{fig:o3_hb} indicate an abrupt change in the physical conditions of the ionized gas between the C1, C2 and C3 regions and their immediate surroundings. 

\subsection{Oxygen abundance and ionization parameters}\label{sec:pyqz}

We now delve deeper into the physics of emission line ratios, and compute oxygen-abundances and ionization parameters for each 353 spaxels with S/N(H$\alpha$;H$\beta$)$\geq$5 and S/N([{\Oiii}];[{\Nii}];[{\Sii}])$\geq$2. The concept and the method employed here follow \cite{Dopita13b}. We use the \textsc{pyqz} v0.6.0 \textsc{python} module to compute, for each spaxel, the value of the oxygen abundance 12+$\log$(O/H) and the ionization parameter $\log$(q) (with q in cm s$^{-1}$). The code relies on a series of grids of photoionization models computed with \textsc{mappings iv} - the most recent embodiment of the \textsc{mappings} code \citep[][]{Dopita82,Binette82,Binette85,Sutherland93, Groves04, Allen08} - which for a distinct set of line ratio diagrams allow to disentangle the values of 12+$\log$(O/H) and $\log$(q) over a large parameter space ($5 - 0.05~Z_{\odot}$ and $6.5 \leq \log (\rmn{q}) \leq 8.5$). Among other updates, \textsc{mappings iv} allows for a non-Maxwellian distribution for the electron energies in the form of a $\kappa$-distribution \citep{Nicholls12, Nicholls13, Dopita13a}. Here, we adopt $\kappa$=20, but we note that this choice does not significantly affect our analysis, given the limited influence of a $\kappa$ distribution on the intensity of strong emission lines. 

We note that for this work we have upgraded \textsc{pyqz} from v0.4.0 publicly released by \cite{Dopita13a} to v0.6.0. In this latest version, \textsc{pyqz} can fully propagate errors in the line flux measurements through to the estimation of the abundance and ionization parameters \cite[but still relies on the same \textsc{mappings iv} models of][]{Dopita13a}. The new \textsc{pyqz} v0.6.0 will be publicly released in the near future along with updated \textsc{mappings iv} models (Sutherland et al., in prep.), and is available \emph{on-demand} until then. For completeness, we describe in Appendix~\ref{app:pyqz} how observational errors are propagated in this new version of \textsc{pyqz} and how the final 1-sigma uncertainty level on the 12+$\log$(O/H) and $\log$(q) values is computed, namely via the propagation of the full probability density function associated with each emission line measurement. 

In total, eight diagnostic grids are available in \textsc{pyqz} to compute the abundance and ionization parameter of a given spectrum. In principle, each of these grids can provide an estimate of 12+$\log$(O/H) and $\log$(q), which can then be combined to provide a global estimate. In practice, and for the present case, we only used the diagnostics not involving [{\Oii}] because these emission lines are subject to flat-fielding issues (see Section~\ref{sec:wifes_reduc}). In addition, an inherent twist in the model grid over the region of interest for HCG~91c renders the diagram $\log$~$[${\Nii}$]$/H$\alpha$ \emph{vs} $\log$~[{\Oiii}]/H$\beta$ unusable in the present case. Ultimately, we are left with just two suitable diagnostic grids: $\log$~$[${\Nii}$]$/[{\Sii}] \emph{vs}  $\log$~[{\Oiii}]/H$\beta$ and $\log$~$[${\Nii}$]$/[{\Sii}] \emph{vs}  $\log$~[{\Oiii}]/[{\Sii}].

\begin{figure}
\centerline{\includegraphics[scale=0.4]{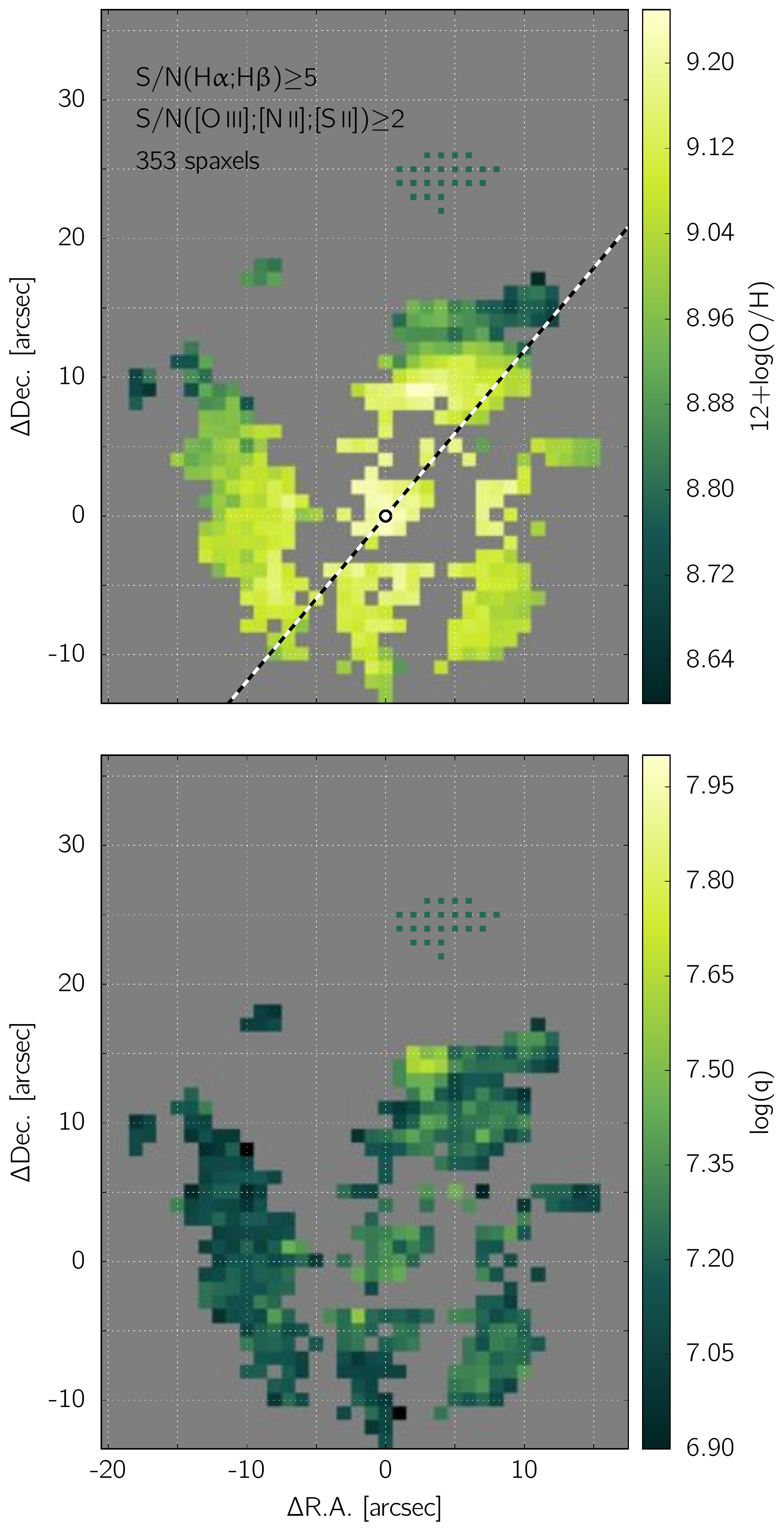}}
\caption{Gas phase oxygen abundance (top) and ionization parameter (bottom) measured with the \textsc{pyqz} \textsc{python} module for all spaxels with S/N(H$\alpha$;H$\beta$)$\geq$5 and S/N([{\Oiii}];[{\Nii}];[{\Sii}])$\geq$2. In the top panel, the dashed line traces the major axis of HCG~91c with a P.A. of 40 degrees West-of-North (see Section~\ref{sec:wifes_kin}).}\label{fig:qz_map}
\end{figure}

The resulting 12+$\log$(O/H) and $\log$(q) maps for HCG~91c are shown in Figure~\ref{fig:qz_map}. As mentioned previously, the overall decrease of the [{\Oiii}] flux in the inner regions of HCG~91c is associated with increasing oxygen abundances. The peak oxygen abundance at the galaxy center is 12+$\log$(O/H)$\approx$9.25. By comparison, the oxygen abundance of the R1 region is found to be 12+$\log$(O/H)$\approx$8.80. The oxygen abundance of the C1, C2 and C3 regions is of the order of 8.7-8.9. The ionization parameter map is uniform throughout the entire galaxy ($7.10 \la \log (\rmn{q}) \la 7.35$). The C1 region is a notable and the only significant departure from the mean with $\log$(q)$\approx$7.65. 

\begin{figure}
\centerline{\includegraphics[scale=0.4]{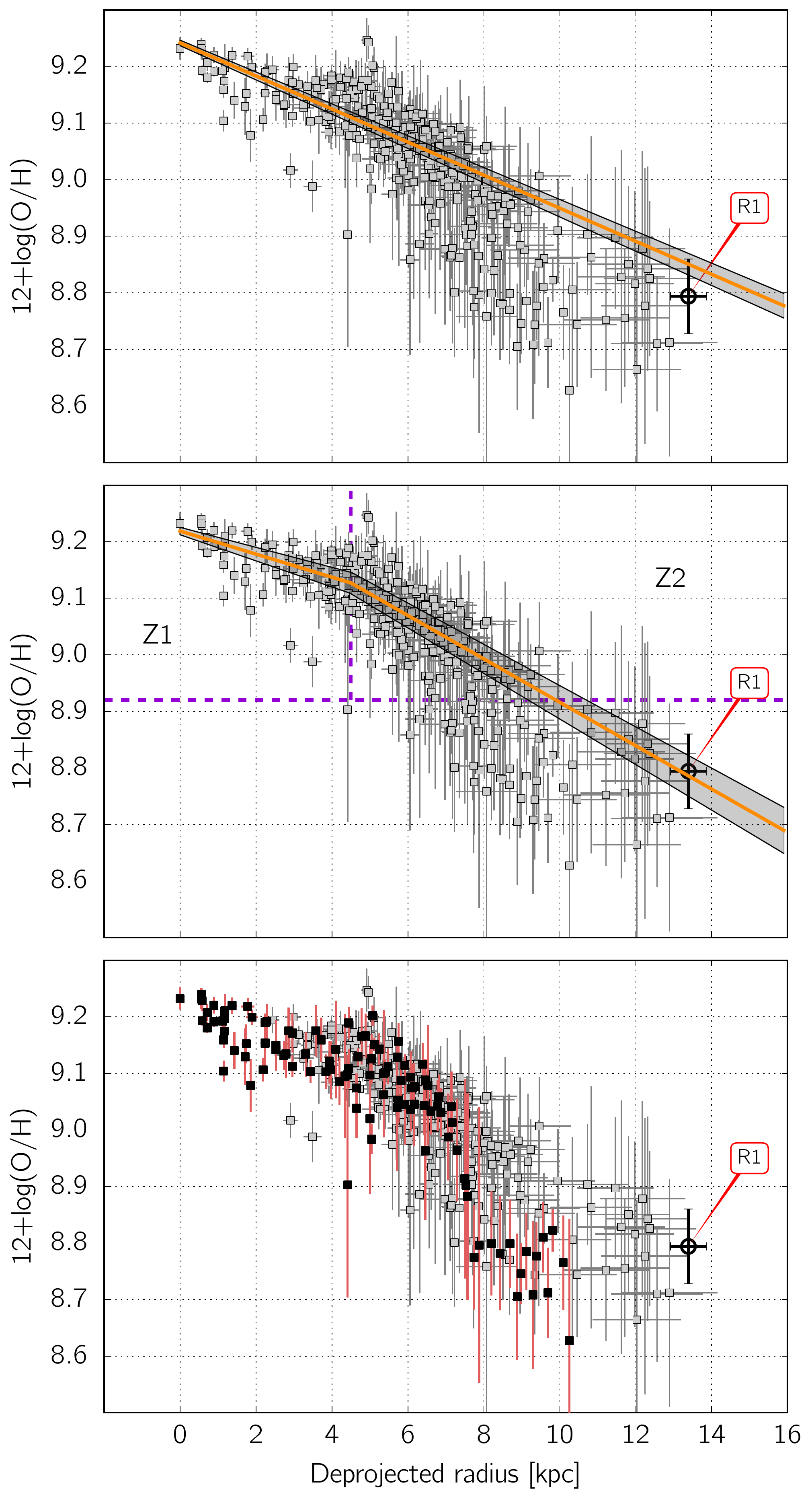}}
\caption{Oxygen abundance gradient in HCG~91c. Each 353 spaxels for which 12+$\log$(O/H) could be measured (see Figure~\ref{fig:qz_map}) are shown individually with their associated 1-sigma errors. The galaxy center, taken to be the peak of the H$\alpha$ emission line flux map, is at the position [0,0] (see Figure~\ref{fig:ha_map}). A position angle P.A. = 40$^{\circ}$ and an ellipticity $b/a$=0.8 are assumed to deproject the radius of each spaxel within the disk of HCG~91c. The horizontal error bars correspond to a measurement error of 10 per cent of the ellipticity. The vertical error bars are the error computed by \textsc{pyqz} v0.6.0. The linear best fit (and associated 1-sigma error) to all the spaxels is shown in the top panel. A broken linear fit, better matching the outer regions of the galaxy, is shown in the middle panel. In the bottom panel, the spaxels located within 3 arcsec from the galaxy's major axis (shown in Figure~\ref{fig:qz_map}) are drawn in black with their associated errors in red. They reveal the abrupt drop of $\sim$0.2 dex in oxygen abundance at $\sim$7.5 kpc to the North-West of the galaxy center: the location of the C1, C2 and C3 star forming regions.}\label{fig:z_broken_grad}
\end{figure}

In Figure~\ref{fig:z_broken_grad}, we construct the abundance gradient of HCG~91c as a function of the deprojected radius. Every 353 spaxels for which we derived an oxygen abundance are shown individually. The vertical error bar associated to each measurement  corresponds to the error computed by \textsc{pyqz} v0.6.0. To deproject the position of each spaxel in the disk of HCG~91c, we assume an ellipticity $b/a$=0.8$\pm$0.08 (measured manually from the r$_{\rmn{P1}}$ image of HCG~91c from Pan-STARRS), and a position angle P.A.=40$^{\circ}$ West-of-North (measured from the rotation map of HCG~91c, see Section~\ref{sec:wifes_kin}). This P.A. is in disagreement with the P.A. available in NED based on near-IR images from 2MASS \citep{Skrutskie06} of 5$^{\circ}$ West-of-North, but is in good agreement with the P.A. derived from optical images from the \emph{Surface Photometry Catalogue of the ESO-Uppsala Galaxies} \citep{Lauberts89} of 46$^{\circ}$ West-of-North. The horizontal error bar associated with each measurement correspond to an estimated 10 per cent measurement error on the ellipticity. We assumed the center of the galaxy to be located at the position [0,0], coincident with the peak H$\alpha$ emission, which given the seeing conditions is also consistent with the kinematic center of the galaxy, as described in the next Section.

As illustrated in the top panel of Figure~\ref{fig:z_broken_grad}, a linear fit to the full set of data points fails to properly match the outer regions of the disk. It is clear from Figure~\ref{fig:qz_map} that our sampling of the outer regions of the disk (where 12+$\log$(O/H)$\leq$8.92) is poor, and largely composed of the C1, C2 and C3 regions. Hence, the global gradient fit is largely influenced by the spaxels in the inner regions of HCG~91c. In the middle panel of Figure~\ref{fig:z_broken_grad}, we use a linear \emph{broken} fit to better reproduce the trend in the oxygen abundance throughout HCG91c. The fit is broken at 4.5~kpc from the galaxy center - the approximate radius at which the oxygen abundance gradient appears to steepen. An \emph{inner gradient} is derived from spaxels in the Zone 1 (Z1; radius$<$4.5kpc, 12+$\log$(O/H)$>$8.92), while an \emph{outer gradient} is derived from spaxels in the Zone 2 (Z2; radius$\geq$4.5kpc, 12+$\log$(O/H)$>$8.92), and forced to match the inner gradient at 4.5 kpc. The respective slopes of the different gradients are compiled in Table~\ref{table:gradients}. All oxygen abundance gradients were derived using the Orthogonal Distance Regression routines inside the \textsc{scipy} module in \textsc{python}\footnote{http://docs.scipy.org/doc/scipy/reference/odr.html, accessed on 2015 February 12}.

A series of spaxels beyond 7 kpc from the galaxy center remain below the best fit outer gradient by $\sim$0.15 dex. In the bottom panel of Figure~\ref{fig:z_broken_grad}, we show in black the spaxels located within 3 arcsec from the galaxy's major axis visible in Figure~\ref{fig:qz_map}. These spaxels trace an abrupt drop in the oxygen abundance of $\sim$0.2 dex at 7 kpc from the galaxy center over a distance of 1 kpc = 2 WiFeS spaxels. This sharp oxygen abundance decrease is essentially unresolved (spatially) in our data set, so that its true nature (physical discontinuity in the oxygen abundance or sharp but continuous decrease) remains uncertain. It is however clear that the intensity of the oxygen abundance drop at 7 kpc is inconsistent with any linear gradient extrapolated from the inner regions of HCG~91c.

We deliberately use only spaxels in the zone Z2 to derive the best-fit linear outer gradient, as we detect spaxels with 12+$\log$(O/H)$>$8.92 for (almost) all azimuths beyond a radius of 4.5 kpc. Our sampling of less enriched gas is much less uniform, so that including all the spaxels with 12+$\log$(O/H)$<$8.92 in the outer gradient fit would wrongly bias it towards the C1, C2 and C3 regions of which the abundances (as illustrated in the bottom panel of Figure~\ref{fig:z_broken_grad}) are clearly inconsistent with a linear gradient. Although the outer gradient is derived solely from the Z2 spaxels,  we note that its slope nevertheless matches some of the outer regions of HCG~91c, including the R1 region.

\begin{table}
\caption{ Characteristics of the different oxygen abundance gradients presented in Figure~\ref{fig:z_broken_grad}.}\label{table:gradients}
\begin{center}
\begin{tabular}{l c c c}
\hline
\hline
Gradient & Fitted region(s) & Slope& Zero point \\[1ex]
&& [dex kpc$^{-1}$] &\\
\hline
global fit & all & -0.029$\pm$0.001 &  9.242$\pm$0.005 \\[3ex]
inner fit & Z1 & -0.020$\pm$0.003 &  9.219$\pm$0.007 \\[3ex]
outer fit & Z2 & -0.038$\pm$0.002 &  9.300$\pm$0.011 \\[1ex]
\hline
\end{tabular}
\end{center}
\end{table} 

\begin{figure}
\centerline{\includegraphics[scale=0.35]{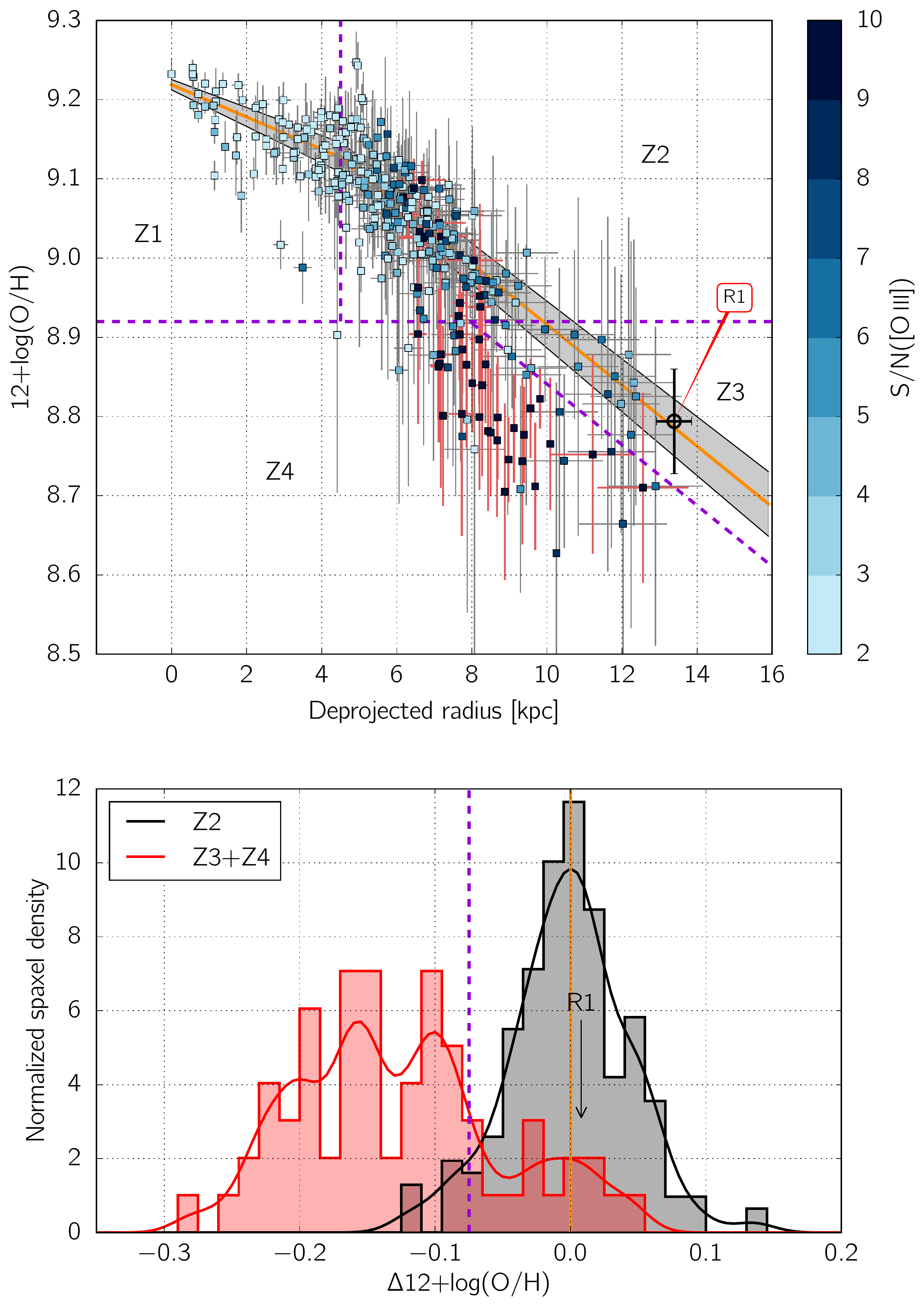}}
\caption{Top: same as the middle panel of Figure~\ref{fig:z_broken_grad}, but with the individual spaxels colored as a function of S/N([{\Oiii}]). The diagram is divided in four gradient zones (labelled Z1 to Z4) based on the overall trend morphology. For clarity, errors for spaxels with S/N([{\Oiii}])$\geq$9 are shown with thicker dark-red lines. Bottom: histogram and associated smooth distribution of the oxygen abundance offset from the best-fit outer-gradient, for spaxels in the Z2 zone (black) and in the Z3+Z4 zones (red). }\label{fig:z_gradient}
\end{figure}

In the top panel of Figure~\ref{fig:z_gradient}, we color-code the different spaxels as a function of their associated S/N([{\Oiii}]. Most spaxels lying in the Zone 4 (Z4; 12+$\log$(O/H)$<$8.92, 12+$\log$(O/H) more than 0.075 dex below the best fit outer gradient based on the Z2 spaxels) are clearly detected, and their errors (shown in dark red for clarity) place them 1-2 sigma below our best-fit outer gradient. Their offset in oxygen abundance is best seen in the Figure's bottom panel, showing the distribution of oxygen abundances for all spaxels in the zones Z2 and Z3+Z4. While some spaxels below 12+$\log$(O/H)=8.92 are consistent with the linear gradient estimate based on the Z2 spaxels out to 13 kpc (i.e. spaxels in the zone Z3), most lie 0.15 - 0.3 dex lower (in the zone Z4).
 
\begin{figure}
\centerline{ \includegraphics[scale=0.4]{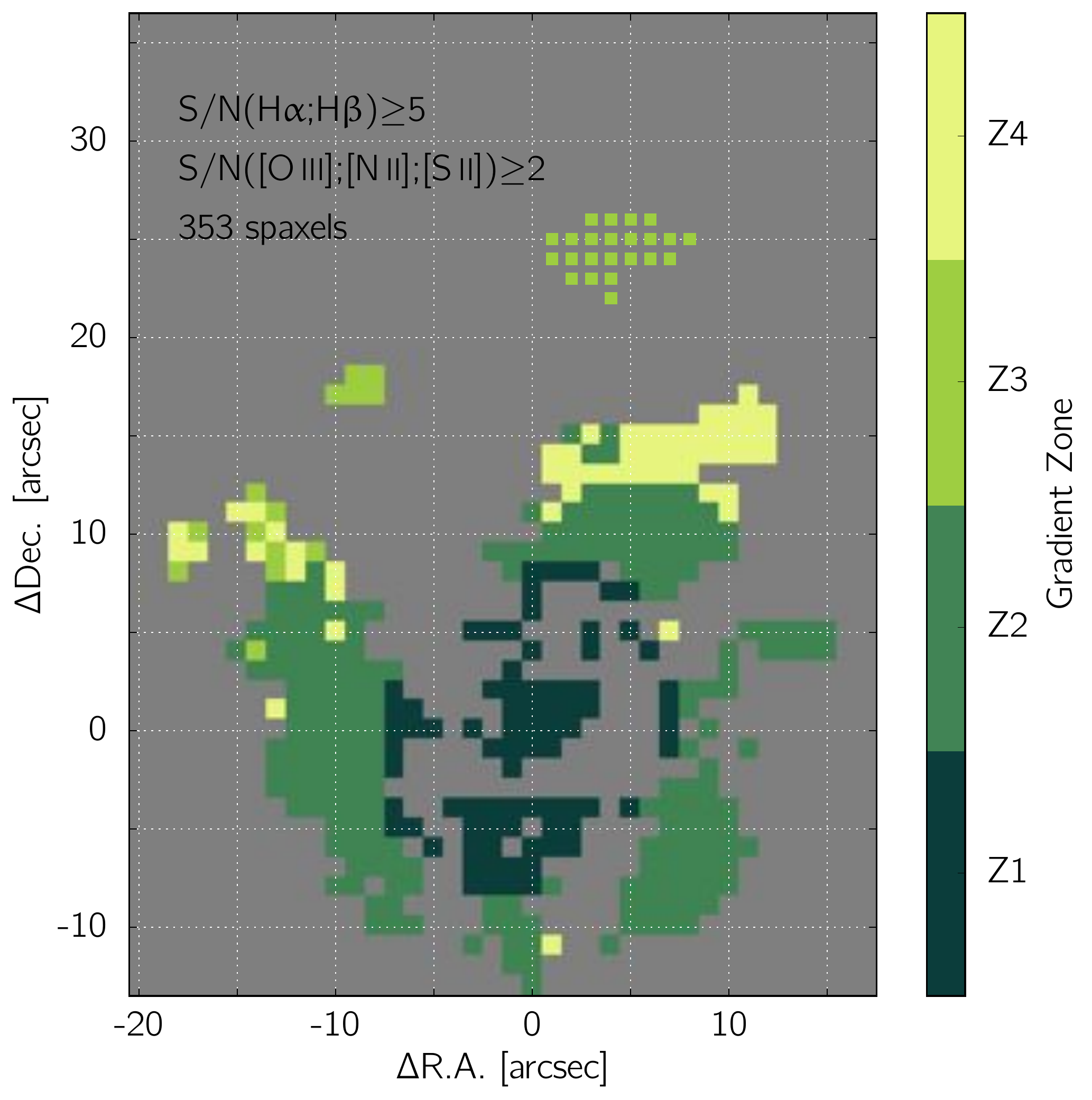}}
\caption{Spatial distribution of the spaxels in the different gradient zones Z1 to Z4 defined in Figure~\ref{fig:z_gradient}.}\label{fig:zone_map}
\end{figure}

Most spaxels in the zone Z4 belong to the C1,C2 and C3 regions, as illustrated in Figure~\ref{fig:zone_map}. It can be noted that as the C1, C2 and C3 regions are located along or near the P.A. of HCG~91c, their deprojected radii are not subject to large uncertainties. The Z3 spaxels are on the other hand all located in the outer regions of the spiral arm extending Northward from the left-hand side of HCG~91c. Especially, the lack of S/N in the [-10;15] area (see Figure~\ref{fig:qz_map}) appears consistent with the lack of detection of Z3 spaxels at a radius of 10-11~kpc.

\subsection{Kinematics of the ionized gas}\label{sec:wifes_kin}

\begin{figure*}
\centerline{\includegraphics[scale=0.45]{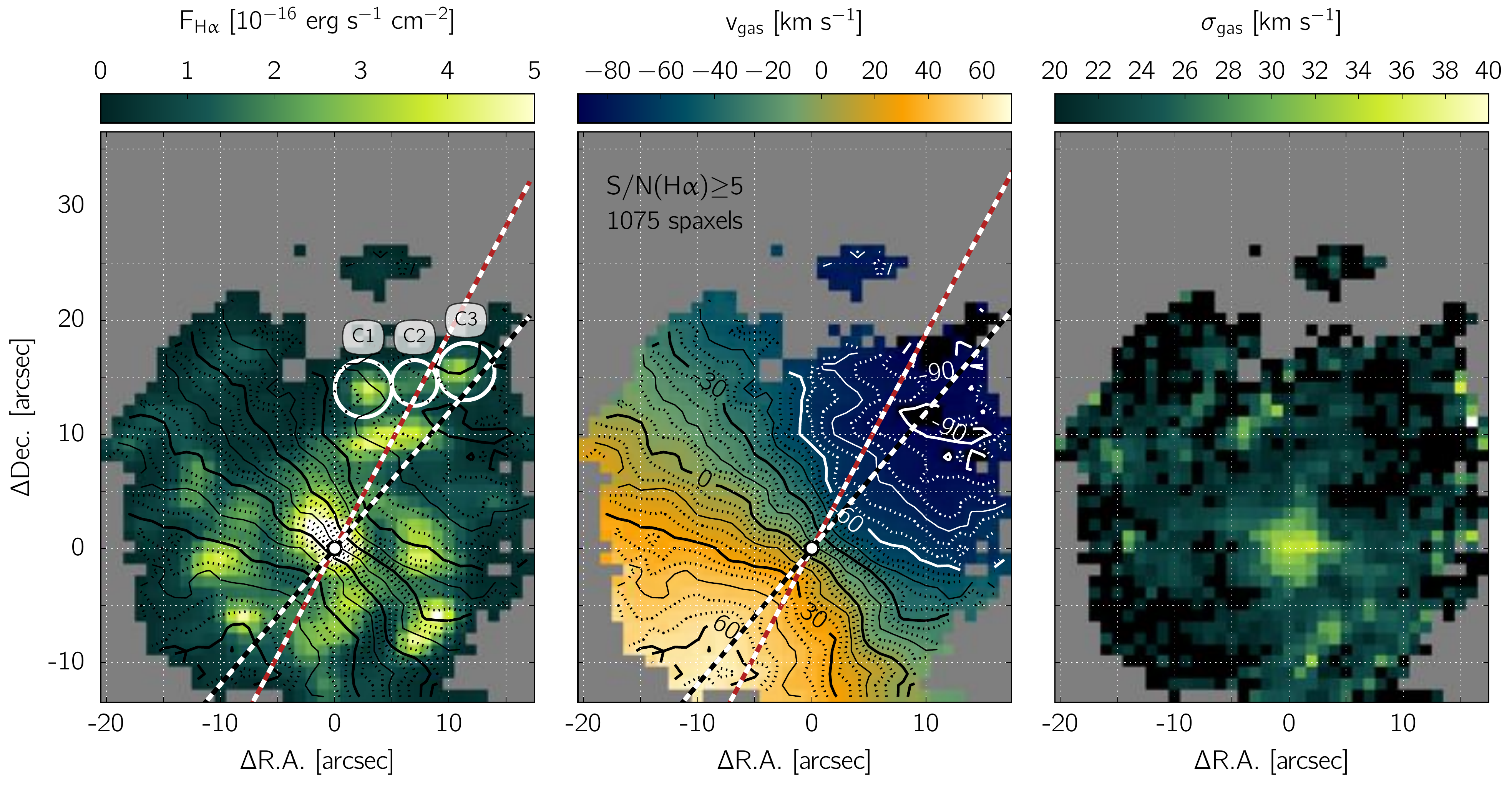}}
\caption{Velocity (center) and velocity dispersion (right) maps of HCG~91c, for all spaxels with S/N(H$\alpha$)$\geq$5. Iso-velocity contours are shown in the middle panel, and are also overlaid over the H$\alpha$ flux map of HCG~91c is the left panel. In the left and middle panel, the center of the galaxy (traced by the peak H$\alpha$ line flux) is marked with a white circle, and the black \& white dashed line traces the principal axis of HCG~91c (P.A.=40$^{\circ}$ West-of-North). The dark-red \& white dashed line is rotated by 12$^{\circ}$ counter-clockwise from the galaxy's major axis, and is designed to pass through the C1-C2-C3 complex of star forming regions. The Position-Velocity diagrams in Figure~\ref{fig:pv} are extracted along these two directions. For the iso-velocity contours, the thick-full lines are spaced by 30 km s$^{-1}$, the thin-full lines are spaced by 15 km s$^{-1}$ and the dashed lines are spaced by 5 km s$^{-1}$.}\label{fig:vsig}
\end{figure*}

In Figure~\ref{fig:vsig}, we show the ionized gas velocity map (center) and velocity dispersion (right) of HCG~91c. In the left panel, we show the H$\alpha$ intensity map overlaid with the iso-velocity contours.  We use the freely available \textsc{python} routine \textsc{fit}\textunderscore\textsc{kinematic}\textunderscore\textsc{pa} from M. Cappellari\footnote{http://www-astro.physics.ox.ac.uk/$\sim$mxc/software/, accessed on 2015 February 12} to measure the P.A. of HCG~91c from its gas kinematics: P.A. = 40 $\pm$ 4 degrees West-of-North. This routine (originally written in \textsc{idl}) was used extensively by \cite{Cappellari07} and \cite{Krajnovic11}, and is described in Appendix C of \cite{Krajnovic06}. The P.A.=40$^{\circ}$ (black \& white dashed line) direction and the galaxy center (white dot) used to construct the abundance gradient of HCG~91c shown in Figure~\ref{fig:z_gradient} are shown in the left and middle panel of Figure~\ref{fig:vsig} for completeness. 

The kinematic signature of the gas in HCG~91c is consistent with regular rotation. The quoted velocities assume a rest frame velocity of 7319 km s$^{-1}$ \citep[z=0.024414;][]{Hickson92}.  We find an overall asymmetry between the redshifted and blueshifted sides that could be reconciled if the rest-frame velocity was reduced by 11 km s$^{-1}$ to 7308 km s$^{-1}$ (as measured by \textsc{fit}\textunderscore\textsc{kinematic}\textunderscore\textsc{pa}). This offset would also help reconcile the optical redshift of HCG~91c with the {\Hi} kinematic of the galaxy measured by the \emph{VLA} (see Section~\ref{sec:group}). 

The velocity dispersion of the ionized gas is throughout the entire disc of HCG~91c ranging from $\sim$20 km s$^{-1}$ to 40 km s$^{-1}$. Accounting for the thermal broadening of the lines \citep[$\sigma_\text{Th}\approx 13 (T/10^4)^{0.5}$ km s$^{-1}$ for the Hydrogen lines, see][]{Osterbrock89}, this corresponds to an intrinsic velocity dispersion range of $\sim$15-38 km s$^{-1}$. The largest velocity dispersions are found towards the galaxy center, and are most certainly influenced by beam smearing, given the seeing conditions during our observations (1.2-1.5 arcsec). No significant increase of the velocity dispersion is detected towards the C1, C2 and C3 clumps.

We detect some deviations from a regular rotation signature at and around the positions [5,15] to [12,12] - the location of the C1, C2 and C3 regions. Specifically, the C1, C2 and C3 regions are redshifted from (i.e. are lagging behind) the regular rotation of the galaxy by 5-10 km s$^{-1}$. We recall that the error associated with the spectral calibration of the red WiFeS datacube is of the order of 0.05{\AA} \citep[see Section~\ref{sec:wifes_reduc} and][]{Childress14a} or 2.3 km s$^{-1}$ at H$\alpha$, so that a 5-10 km s$^{-1}$ offset corresponds to a 2-3 sigma detection. The fact that the kinematic distortion is extended and tracks the bright spaxels distribution of the C1, C2 and C3 regions suggest that the distortion is real, although a greater spectral resolution and finer spatial sampling would be beneficial to confirm its existence. 

\begin{figure}
\centerline{\includegraphics[scale=0.38]{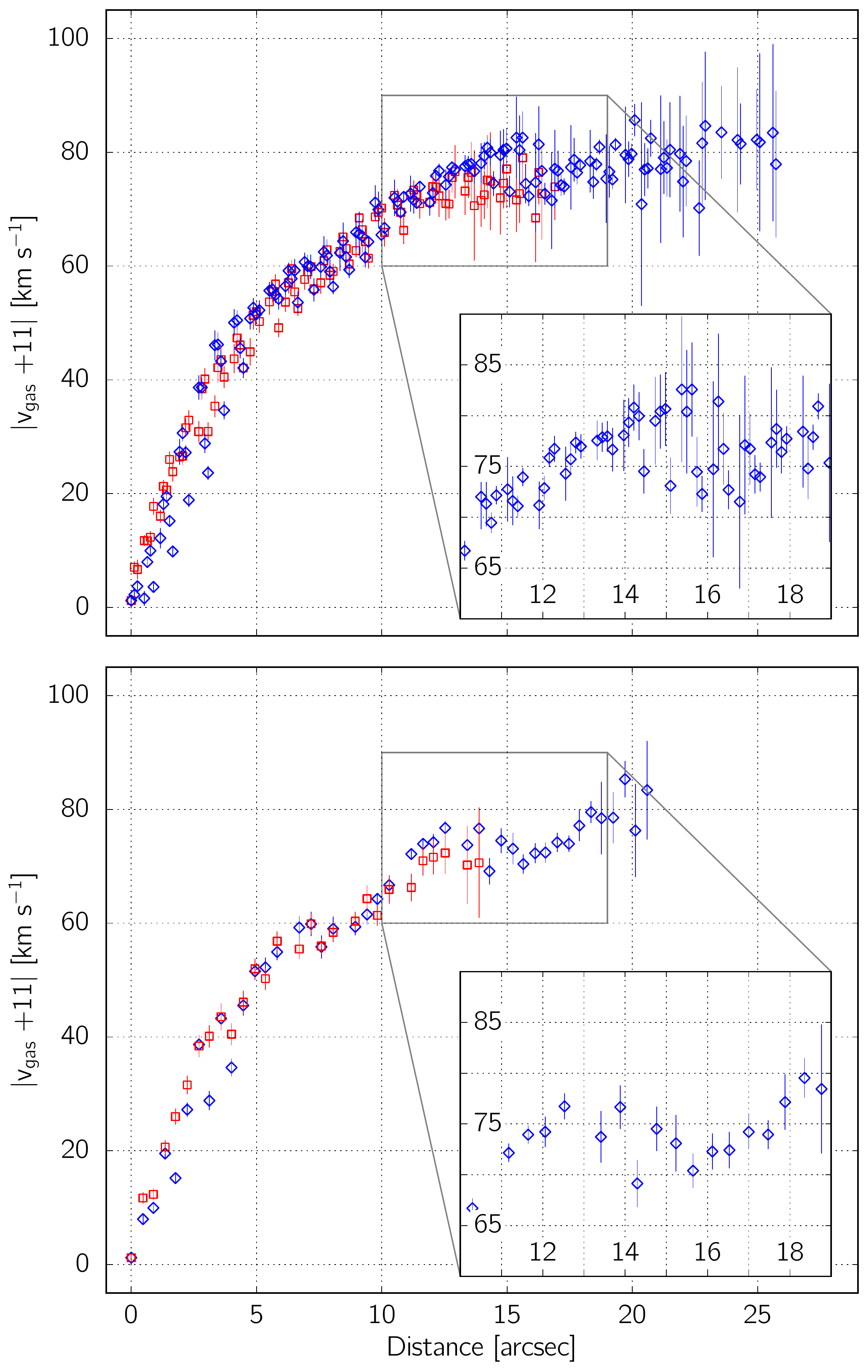}}
\caption{Top: Position-Velocity diagram of the velocity map shown in Figure~\ref{fig:vsig}, extracted along the axis passing through the galaxy center and with P.A. = 40$^{\circ}$ West-of-North (shown with a black \& white dashed line in Figure~\ref{fig:vsig}). Each spaxel within 3 arcsec of the axis and with S/N(H$\alpha$)$\geq$5 is shown as a red square (redshifted side of HCG~91c) or a blue diamond (blueshifted side of HCG~91c). A correction of 11 km s$^{-1}$ was added to the velocity of each spaxel to correct the global kinematic asymmetry detected in the velocity field. Bottom: idem, but for spaxels with S/N(H$\alpha$)$\geq$5 located within 1 arcsec from an axis rotated by 12$^{\circ}$ counter-clockwise from the galaxy's major axis (traced with a dark-red \& white dashed line in Figure~\ref{fig:vsig}).}\label{fig:pv}
\end{figure}

To further quantify the distortions of the velocity map of HCG~91c associated with the C1, C2 and C3 star forming regions, we construct two Position-Velocity  (PV) diagrams. We choose the projection axis to be oriented a) along the galaxy's P.A. and b) along an axis rotated by 12$^{\circ}$ counter-clockwise from the galaxy's major axis. In both cases, we set the zero-point at the galaxy center and add 11 km s$^{-1}$ to the measured velocities to correct the global kinematic asymmetry mentioned previously. The value of 12$^{\circ}$ bears no special significance other than ensuring that the second axis passes through the C1-C2-C3 complex of star forming regions. The resulting PV diagrams are shown in Figure~\ref{fig:pv}. Every spaxel within 3 arcsec and 1 arcsec (respectively) from the two projection axis and with S/N(H$\alpha$)$\geq$5 for which the line velocity can be measured accurately is shown as an individual red square (redshifted side of the galaxy) or a blue diamond (blueshifted side of the galaxy), as a function of its (projected) distance along the PV axis.

Along the major axis of the galaxy, after a linear increase out to $\sim$2~arcsec, the slope of rotation velocity decreases and remains consistent between the red- and blueshifted sides out to $\sim$12~arcsec. While the redshifted gas velocity flattens out at $\sim$75 km s$^{-1}$ beyond 12~arcsec, the blueshifted gas increases up to 80$\pm$3 km s$^{-1}$ beyond 22~arcsec. Assuming an ellipticity for HCG~91c of 0.8$\pm$0.08, we find the absolute rotation velocity of the gas at a radii greater than 11~kpc to be 100$\pm$11 km s$^{-1}$.

A noticeable feature of the top diagram is the kinematic behaviour of the blueshifted gas from 12 to 20~ arcsec. As the magnified inset diagram in Figure~\ref{fig:pv} illustrates, two velocity ``branches'' separated by 5-10 km s$^{-1}$ at the 1-2 sigma level exist in this range. Especially, the ``bottom'' branch corresponds to spaxels in the C3 region, and is the signature of the lag of the C3 star forming region mentioned previously. The velocity lag of the C1, C2 and C3 complex of star forming regions is best seen in the bottom panel of Figure~\ref{fig:pv}, in which the blue velocity curve decrease by $\sim$5-10 km s$^{-1}$ at 16 arcsec from the galaxy center (i.e. at the location of the C2 star forming region).

Spiral galaxies in compact groups have been observed to host a wide range of rotation signatures with sometimes large asymmetries and/or perturbations \citep[see e.g.][]{Rubin91}. As compact groups favor strong gravitational interactions between galaxies, the existence of small perturbations in the velocity field of HCG~91c (located inside a compact group) is in itself not surprising. Of interest however is the fact that the star forming regions with localized lower oxygen abundances are associated with localized kinematic anomalies. This abundance$\leftrightarrow$kinematic connection indicates that both the gas \emph{composition} and \emph{motion} in the C1, C2 and C3 star forming regions are inconsistent with their immediate surroundings.

\subsection{Star formation rate}

\begin{figure}
\centerline{\includegraphics[scale=0.4]{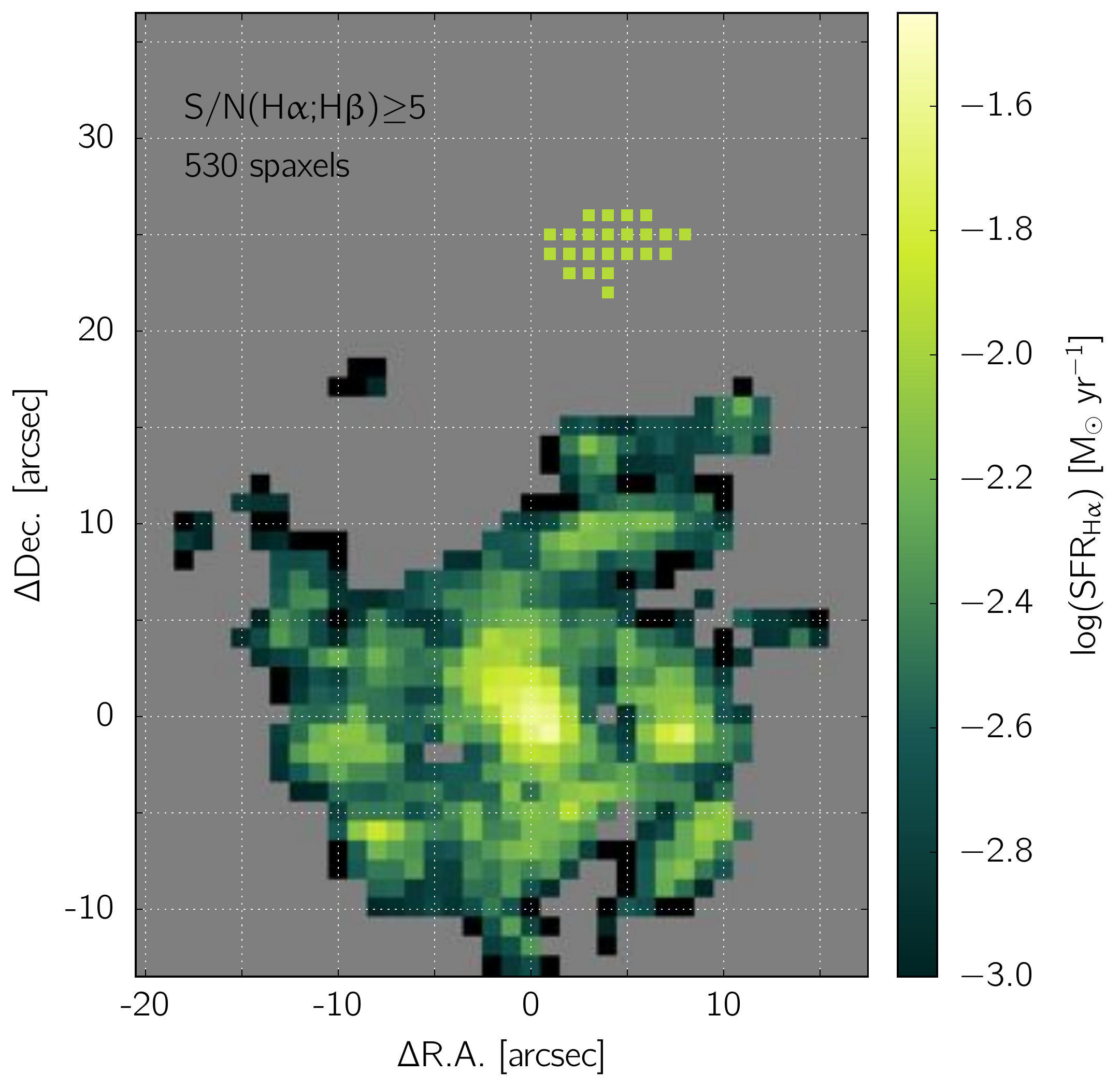}}
\caption{Star formation rate derived from the H$\alpha$ flux, for all spaxel corrected for extragalactic reddening. The galaxy center is the most active region, and the spiral structure is also visible. Localized enhancements of the star formation rate are detected in the C1, C2 and C3 regions.}\label{fig:sfr_map}
\end{figure}

We can convert the H$\alpha$ line flux of each spaxel in our WiFeS mosaic (see Figure~\ref{fig:ha_map}) into a star formation rate (SFR$_{\rmn{H}\alpha}$) following the recipe of \cite{Murphy11} derived using \textsc{starburst99} \citep{Leitherer99} and assuming a \cite{Kroupa01} initial mass function (IMF):
\begin{equation}
\rmn{SFR}_{\rmn{H}\alpha}=5.37\times10^{-42} \frac{L_{\rmn{H}\alpha}}{\rmn{erg s}^{-1}},
\end{equation}
where $L_{\rmn{H}\alpha}$ is the total H$\alpha$ luminosity, derived from our measured de-reddened fluxes per spaxel and given the assumed distance to HCG~91c of 104 Mpc. This recipe results in a SFR 68 per cent of what would be derived using the \cite{Kennicutt98} formula, largely because of differences in the assumed IMF characteristics \citep[][]{Calzetti07,Kennicutt12}. The resulting SFR map of HCG~91c is shown in Figure~\ref{fig:sfr_map}, where only the 530 spaxels with S/N(H$\alpha$; H$\beta$)$\geq$5 and the R1 region that were corrected for extragalactic reddening are shown. We note that our spatial resolution of 0.5~kpc arcsec$^{-1}$ is similar to the scale of the star forming complexes studied by \cite{Murphy11}, so that issues associated with using a ``mean'' H$\alpha$-to-SFR conversion factor on small scales does not apply in our case \citep[see][]{Murphy11, Kennicutt12}.

As expected from the H$\alpha$ emission line map, star formation activity in HCG~91c is most intense in the central region of the galaxy, with additional hot spots along the spiral arms.  The total SFR$_{\rmn{H}\alpha}$ for HCG~91c (summed from all the spaxel with S/N(H$\alpha$; H$\beta$)$\geq$5 and the R1 region) is 2.10$\pm$0.06 M$_{\odot}$ yr$^{-1}$. This value is in excellent agreement with the estimated 2.19 M$_{\odot}$ yr$^{-1}$ from \cite{Bitsakis14} based on a full SED\footnote{Spectral Energy Distribution} fitting of the integrated light of HCG~91c. 

\begin{figure*}
\centerline{\includegraphics[scale=0.35]{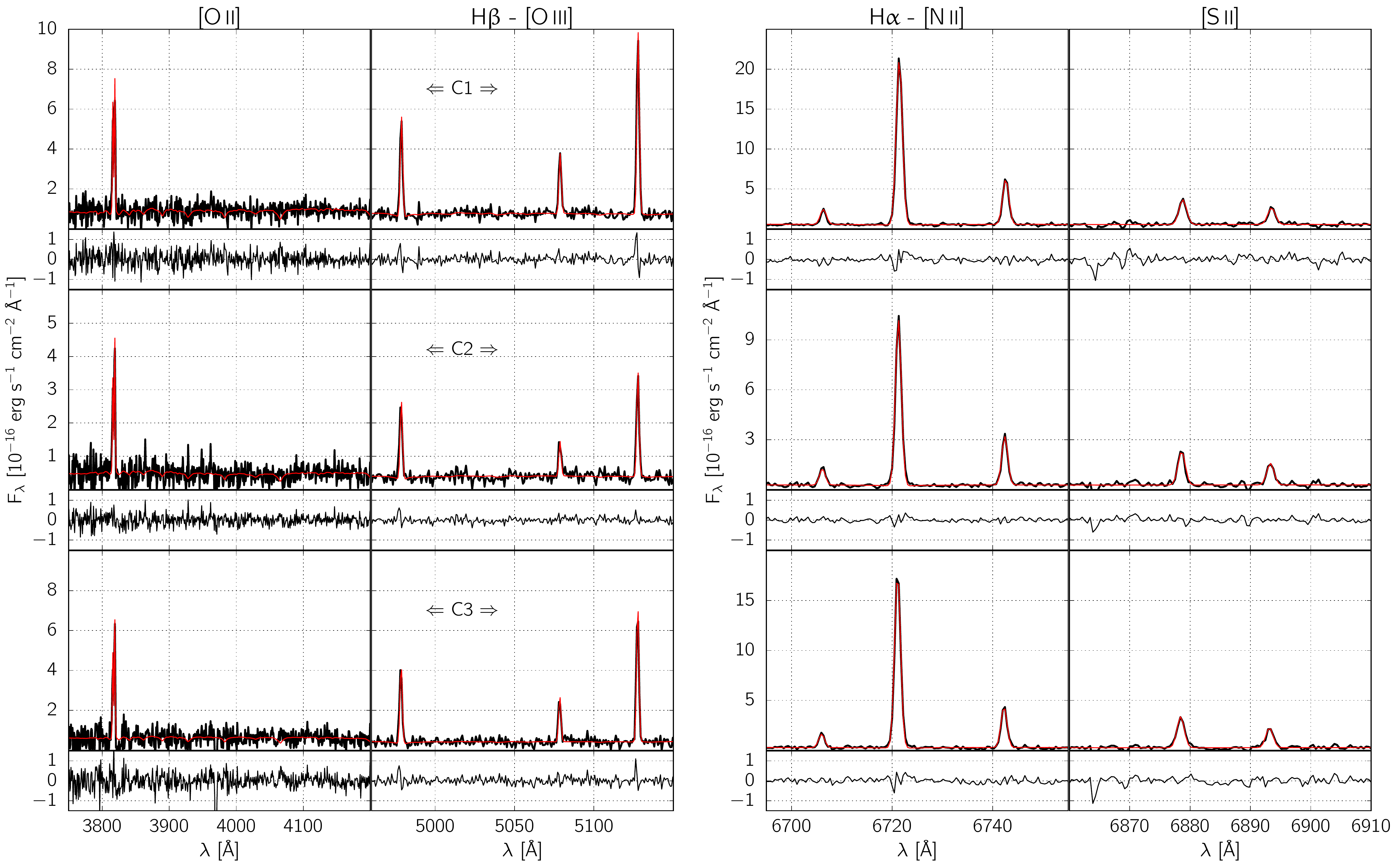}}
\caption{Raw integrated spectra (black) and \textsc{lzifu} fit (red) of the C1, C2 and C3 star forming regions. The dominant emission lines present in each panel are labelled on top of each column for clarity.}\label{fig:spec_fit}
\end{figure*}

Assuming a total stellar mass M$_{*}$=1.86$\times10^{10}$ M$_{\odot}$ for HCG~91c \citep{Bitsakis14}, we obtain a specific star formation rate sSFR = 1.13$\pm$0.05$\times10^{-10}$ yr$^{-1}$. This is comparable to the lowest sSFR measured by \cite{Tzanavaris10} in 21 star forming HCG galaxies from their UV and IR colors, but well within the distribution of sSFR reported by \cite{Plauchu12} for $\sim$50 late-type galaxies (SBc and later) in compact groups using full optical spectrum fitting with stellar population synthesis models. In fact, HCG~91c is extremely consistent with the SFR vs M$_{*}$ and sSFR vs M$_{*}$ relations constructed from the SDSS and GAMA datasets for  $z<0.1$ star forming galaxies by \cite{Lara13}. For reference, we list in Table~\ref{table:clumps} the different characteristics of the C1, C2 and C3 regions, whilst their integrated spectra are presented in Figure~\ref{fig:spec_fit}.

\begin{table}
\caption{Main integrated characteristics of the C1, C2 and C3 star forming regions}\label{table:clumps}
\begin{center}
\begin{tabular}{p{3cm} c c c}
\hline
\hline
Region & C1 & C2 & C3 \\
\hline
\parbox{3cm}{  $<\Delta\rmn{R.A.}>$ \\ $[$ arcsec $]$}               & 3.2  & 7.0 & 10.5 \\[3ex]
\parbox{3cm}{  $<\Delta\rmn{Dec.}>$ \\ $[$ arcsec $]$}               & 13.5  & 13.5 & 15.3 \\[3ex]
\parbox{3cm}{ v $_\text{gas}$ (line of sight) \\ $[$km s$^{-1}$ $]$}               & 7245$\pm$4  & 7236$\pm$3 & 7230$\pm$4 \\[3ex]
\parbox{3cm}{ $\sigma_\text{gas}$ \\ $[$km s$^{-1}$ $]$}               & 21$\pm$2  & 20$\pm$2 & 20$\pm$2 \\[3ex]
\parbox{3cm}{ F(H$\alpha$) \\ $[$10$^{-16}$ erg s$^{-1}$ cm$^{-2}]$}               & 64.0$\pm$8.9  & 23.8$\pm$6.0 & 49.6$\pm$9.0 \\[3ex]
\parbox{3cm}{ SFR$_{\rmn{H}\alpha}$                \\ $[$10$^{-2}$ M$_{\sun}$ yr$^{-1}$$]$}                     &  4.4$\pm$0.6  & 1.7$\pm$0.4 & 3.4$\pm$0.6 \\[3ex]
\parbox{3cm}{ L$_{\rmn{H}\alpha}$                \\ $[$10$^{39}$ erg s$^{-1}$$]$} & 2.1$\pm$0.3  & 0.8$\pm$0.2 & 1.6$\pm$0.3 \\[3ex]
\hline
\end{tabular}
\end{center}
\end{table} 

\section{Discussion: the peculiar star forming regions of HCG~91c }\label{sec:discussion}

By and large, HCG~91c appears as a rather unremarkable star forming spiral galaxy. It hosts regular star formation rates throughout its disc comparable to other similar galaxies in the field, and an overall regular rotation (as traced by the ionized gas). This galaxy also hosts a linear oxygen abundance gradient out to $\sim4.5$ kpc, which then breaks to a steeper linear gradient. The {\Hi} gas distribution around the galaxy traced by the \emph{VLA} is largely undisturbed. The cold dust contours measured by \emph{Herschel/SPIRE} at 250 $\mu$m display a high degree of azimuthal symmetry \citep[][]{Bitsakis14}. \cite{Mendes03} report that HCG~91c is an outlier with respect to the B-band Tully-Fisher relation, in that its absolute B-band magnitude M$_{\rmn{B}}$ is 2 mag brighter than expected from its rotation velocity, and concluded \citep[also based on the two distinct kinematic components reported by][]{Amram03} that HCG~91c may be the result of a merger. We disagree with this picture after finding no evidence for the existence of multiple emission line component in our WiFeS dataset. Furthermore, the clear and regular spiral structure detected with WiFeS and Pan-STARRS strongly argues against the merger scenario, and so is the regular photometric profile of the disk \citep[already noted by][]{Mendes03}.

Altogether, these characteristics suggest that HCG~91c has not (yet) strongly interacted gravitationally with the other galaxy members of HCG~91.The presence of faint, extended {\Hi} gas to the North-West of the main {\Hi} reservoir (see Figure~\ref{fig:dss+VLA}) suggests that HCG~91c may be caught in the very early stage of its interaction with the group. We note that the iso-intensity contours tracing the optical extent of the stellar disk of HCG~91c (see Figure~\ref{fig:panstarrs}) reveal a sharp intensity drop to the South-East, but a smoother and more irregular boundary to the West \& North-West, in the direction of the kinematic asymmetries and {\Hi} clumps. 

Beside these large scale features, we found abundance and kinematic \emph{anomalies} (at the 1-2 sigma level) at the location of the C1, C2 and C3 star forming regions. Our emission line analysis reveals an abrupt change in the physical condition of the ionized gas between these star forming regions and their immediate surroundings. They are comparatively more metal poor by 0.15 dex, and the C1 region hosts the largest value of the ionization parameter in the disk of HCG~91c with $\log$(q)$\approx$7.65. Such large values of $\log$(q) can be explained by a very young age of the associated star forming region ($<$0.5 Myr), a low pressure environment (log(P/$k$)$\approx$4-5 cm$^{-3}$ K), or both \citep[][]{Dopita06}. We also find evidence that the gas in the C1, C2 and C3 regions may be lagging behind the rotation of the disk by 5-10 km s$^{-1}$.

The lack of S/N is our WiFeS observations beyond 7 kpc from the galaxy center hinders us from assessing the detailed behaviour of the oxygen abundance gradient at these large radii. It is clear that the oxygen abundance of the C1, C2 and C3 regions are clearly inconsistent with a linear gradient extrapolated from the inner regions of the galaxy (see Figure~\ref{fig:z_broken_grad}). On the other hand, the oxygen abundances along the spiral arm extending Northward from the top-left of HCG~91c are found to be consistent with the extrapolated linear oxygen abundance gradient. This suggests that the oxygen abundance drop associated with the C1, C2 and C3 star forming regions is localized to these locations, rather than indicative of a uniform drop in the abundance gradient for all azimuths. Determining whether the C1, C2 and C3 star forming regions are unique, or whether more star forming regions in the outer regions of HCG~91c display similar (localized) abrupt decrease in oxygen abundances will require deeper follow-up observations (see Section~\ref{sec:MUSE}).

\subsection{Origin of the fuel for star formation}

How can we explain these localized, anomalous C1, C2 and C3 star forming regions ? Their comparatively lower oxygen abundance (i.e. the presence of less enriched gas) leads us to formulate three possible scenarios to explain their origin: 
\begin{itemize}
\item [(a)] the accretion of a smaller, near-by satellite system,
\item [(b)] gas inflow onto the disk of HCG~91c from the galaxy halo, and 
\item [(c)] the infall and collapse of pre-existing neutral gas clouds at the disk-halo interface of HCG~91c. 
\end{itemize}
For clarity, we discuss each of these scenarios separately. 

\subsubsection{Satellite accretion}

The accretion of a near-by dwarf galaxy could possibly explain the lower metallicity of the C1, C2 and C3 regions. Lower metallicity dwarf galaxies have been detected around the Milky Way and M31, and the presence of large scale stellar streams implies the ongoing accretion and tidal stretching of some of these systems \citep[e.g.][]{Belokurov06, Richardson11}. The apparent on-sky alignment of the C1, C2 and C3 regions (see Figure~\ref{fig:panstarrs}) is certainly suggestive of a \emph{physical} connection. Such a physical connection could then be explained by a tidally disrupted structure. However, the gas kinematic implies difficulties for this scenario. The observed lag of 5-10 km s$^{-1}$ is small compared to typical infall velocities of streams in the halo of M31 and around the Milky Way. Although not impossible, this scenario would require a very specific set of circumstances between the galaxy's orientation, the accretion trajectory, and our point-of-view to result in the small velocity lag of the C1, C2 and C3 star forming regions.

\subsubsection{Gas inflow from the galaxy halo} 

The spatial and kinematic structure of the {\Hi} gas associated with HCG~91c certainly support the idea that this galaxy's halo (still) contains a gravitationally-bound gas reserve. Under these circumstances, the lower oxygen abundance of the C1, C2, and C3 regions could be explained if less enriched and unstructured material from the halo is in-falling onto the disk, and fuelling localized star formation activity. The small velocity lag and the low velocity dispersion of the gas in the C1, C2 and C3 regions would suggest a slow infall velocity. Yet, the lack of any detectable enhancement of the velocity dispersion in the C1, C2 and C3 regions compared to other star forming regions in the disk of HCG~91c remains puzzling. The exact triggering mechanism for the gas infall also remains an open question, but the presence of near-by galaxies and their gravitational fields appear as a likely source of perturbations.

\subsubsection{Infalling and collapsing gas clouds at the disk-halo interface}

In the halo of the Milky Way, there exists a complex mix of gas in different phases, including molecular and neutral hydrogen \citep[][]{Putman12}. Part of this gas is located in a series of compact clouds with a wide range of kinematics \citep{Saul12}. Some of these clouds have measured velocities largely inconsistent with galaxy rotation, and are usually refereed to as \emph{High Velocity Clouds} \citep[HVCs, see e.g.][]{Wakker97}. By comparison, \emph{Intermediate Velocity Clouds}  \citep[IVCs,][]{Wakker01} also have velocities inconsistent with galactic rotation, but less so than HVCs. In fact, there exists a wide range of {\Hi} cloud characteristics in the Milky Way's halo, some of which are found to be co-rotating with the disk \citep[see][and references therein]{Kalberla09}. There also exists diffuse {\Hi} gas in the halo of the Milky Way which displays a vertical lag in its rotation velocity with respect to the disk of the order of 15 km s$^{-1}$ kpc$^{-1}$ \citep{Marasco11}. 

Resolving the structure of the {\Hi} gas in other galaxies is observationally challenging. In recent years, the HALOGAS survey \citep{Heald11} has revealed the presence of an extended {\Hi} disk around edge-on, near-by galaxies \citep{Zschaechner11,Zschaechner12,Gentile13}. These observations complement older detection of gaseous halos, for example around NGC~891 by \cite{Oosterloo07}. Most of these detections revealed a vertical lag of the {\Hi} rotation speed above the galaxy disks of the same order of magnitude than in the Milky Way \citep[but see also][]{Kamphuis13}.

Given the star forming, spiral nature of HCG~91c, it does not appear unreasonable to assume that it possess a multi-phase, complex gaseous halo similar to that of the Milky Way. Especially, although we do not know its detailed structure, the \emph{VLA} observations clearly reveal a large rotating {\Hi} reservoir. Under these circumstances, could the C1, C2 and C3 star forming regions have resulted from the infall and subsequent collapse of pre-existing gas clouds at the disk-halo interface of HCG~91c ? The compact nature of these star forming regions could be naturally explained by a ``localised cloud'' origin. Their velocity lag of 5-10 km s$^{-1}$ would suggest that these star forming regions are in fact not located within the disk of HCG~91c, but 300-700 pc above it, if one assumes a vertical rotational velocity lag of 15 km s$^{-1}$ kpc$^{-1}$ \citep{Marasco11}. Of course, the face-on nature of HCG~91c makes it impossible to directly measure any vertical offset for these star forming regions. However, we note that the associated reddening is lower in the C1, C2 and C3 regions  (A$_{\rmn{V}}\leq0.9$, see Figure~\ref{fig:av}) than in any other star forming regions in the spiral arms of HCG~91c (A$_{\rmn{V}}\geq1.2$) at the 1-sigma level. A lower reddening value may be the result of the location of the C1, C2 and C3 regions above the main disk (and dust) of HCG~91c.

\subsection{Star formation triggering mechanism}

Clearly, we cannot firmly rule out any of the above scenarios invoked to explain the origin of the anomalous star forming regions C1, C2, and C3 in HCG~91c. At this stage, we favor the idea that these star forming regions originated in the infall and subsequent collapse of pre-existing gas clouds at the disk-halo interface. Indeed, collapsing pre-existing gas clouds at the disk-halo interface could naturally explain the different properties of the anomalous star forming regions (velocity lag, lower oxygen abundance, compactness). 

The precise mechanism able to trigger the collapse of neutral gas clouds at the disk-halo interface of HCG~91c (and the subsequent rapid formation of molecular gas to fuel star formation) remains undefined. In the Milky Way, molecular hydrogen was detected in several IVCs \citep{Richter03, Wakker06}, and it has been proposed that neutral hydrogen is compressed into molecular hydrogen as gas clouds fall onto the galaxy disk and get compressed via ram pressure stripping \citep{Weiss99,Gillmon06,Roehser14}. The C1, C2 and C3 star forming regions may have resulted from a boosted version of this mechanism, where the initial infall of neutral gas clouds onto the galaxy disk is being triggered by large scale gravitational perturbations from near-by galaxies in an harassment-like process \citep{Moore98}. The existence of tidal perturbations to the North-West of HCG~91c (supported by the existence of a possible {\Hi} tail and the comparatively complex edge of the stellar disk to the North-West of HCG~91c) may have given rise to local tidal shears or compressive tides \citep[][]{Renaud08,Renaud09}, triggering the collapse of the clouds. Theoretically, \cite{Renaud14} observed the effect of compressive turbulence and how it can lead to starburst events in interacting galaxies. This mechanism is already active during the early phase of galaxy interactions, and could therefore be active in HCG~91c. In the same simulations, large scale gas flows across a galaxy's disk only occur at later stages of the interaction (i.e. at the second closest approach). We note that the initial star formation triggered by compressive turbulence does not appear to require large velocity dispersions ($\sigma<$20 km s$^{-1}$), which would be consistent with our WiFeS observations.

Alternatively, HCG~91c is located $\sim$30 arcsec = 15 kpc (on sky) to the North-East of the extended tidal tail of HCG~91a (see Figure~\ref{fig:dss+VLA}).  HCG~91c's {\Hi} envelope also suggest that the galaxy may have previously interacted with HCG~91b (as indicated by the presence of a possible {\Hi} bridge between the two galaxies). Ram-pressure from either interaction may have shocked and compressed HCG~91c's halo, leading to the collapse of the C1, C2 and C3 star forming regions.  Kinematically, HCG~91c and the tidal tail from HCG~91a are offset by $\sim$250 km s$^{-1}$, but given their spatial location, it is possible that HCG~91c is currently interacting/colliding with this extended tidal structure. Such a collision is impossible to firmly rule out without X-ray observations of the group which may reveal the presence of hot, shocked gas resulting from the interaction. In any case, the relatively undisturbed {\Hi} envelope of HCG~91c and the continuous structure of the tidal tail stemming from HCG~91a detected by the \emph{VLA} would suggest that their interaction is at an early stage.

\subsection{Constraining the exact nature of the C1, C2 and C3 regions}\label{sec:MUSE}

Our WiFeS observations, combined to \emph{VLA} and Pan-STARRS datasets, have revealed the anomalous oxygen abundance of three compact star forming regions in the disk of HCG~91c, and showed that this galaxy is only just beginning its interaction with the compact group HCG~91. From the ionized gas physical characteristics and kinematics, we found that infalling (and subsequently collapsing) gas clouds at the disk-halo interface could naturally explain the anomalous nature of the C1, C2 and C3 regions. However, one should keep in mind that our different pieces of evidence for the anomalous nature of the C1, C2 and C3 star forming regions (velocity and abundance offset, lower extinction) are all at the 1-2 sigma level.

We also lack critical pieces of information: for example, a deeper understanding of the state of the gas (i.e. pressure, density, temperature) in these specific star forming regions, as well as a better characterization of the underlying stellar population. Additional insight on the stellar population associated with the C1, C2 and C3 regions is especially critical to test their possible origins. For example, collapsing halo gas clouds with masses ranging from 10$^3$-10$^5$ M$_{\odot}$ \citep[see][and references therein]{Putman12} do not offer a sustainable source of fuel and cannot host a significant population of old stars, of which the detection would be a very strong argument \emph{against} this scenario. Unfortunately, the S/N in the continuum of our observations does not allow us to extract meaningful information on the underlying stellar population in HCG~91c (see Figure~\ref{fig:spec_fit}). 

The inherent compact aspect of the C1, C2 and C3 regions would also benefit from a finer spatial sampling to reveal their precise structural extent and possible physical connections. Finally, the disk of HCG~91c extends beyond the field-of-view of our WiFeS observations, and may contain additional anomalous star forming regions. To address these questions, we have been awarded 4.0 hr of Science Verification time on MUSE \citep{Bacon10}, the new integral field spectrograph on the Yepun telescope (unit 4 of the \emph{VLT}) at Paranal in Chile (P.I.: F.P.A.~Vogt, P.Id.: 60.A-9317[A]), which will be the subject of a separate article.

We have shown that overall, HCG~91c is a largely undisturbed (yet) star forming spiral galaxy. Hence, HCG~91c offers us a window on the early stage of galaxy interactions in a compact group, and possibly on the early phase of galaxy pre-processing in these environments. The presence of localized star forming regions with lower oxygen abundances indicate that lower-metallicity gas is being brought from the outer regions of the disk or the halo to the inner regions of the galaxy. Hence, this gas may eventually contribute to the flattening of the oxygen abundance gradient in the system. Strong interactions of galaxies in pairs have been observed to flatten the metallicity gradient in these system through large scale gas flows \citep{Kewley10,Rupke10}. Our WiFeS observations of HCG 91c suggest that prior to large scale gas flows induced by strong gravitational perturbation, some gas mixing can be occurring at the level of individual star forming regions as a result of galaxy harassment and longer range gravitational interactions.

Evaluating the ubiquity (or not) of collapsing gas clouds at the disk-halo interface of galaxies as an early consequence of galaxy harassment will require additional observations of a statistical sample of galaxies. Existing or upcoming IFU surveys such as SAMI or MaNGA may provide such a statistical sample probing a wide range of environment densities. However, with a respective spectral resolution of R=4500 and R=2000 and a spatial sampling $\geq$ 1 kpc, finding sub-kpc star forming regions with $\sim$0.15 dex oxygen abundances offsets will certainly prove challenging for these surveys. Rather than ``direct'' detections, the SAMI and MaNGA surveys may instead provide ``tentative'' detections of anomalous star forming regions in largely unperturbed star forming disk galaxies. These objects would form an ideal sample (spanning a wide range of environments) for dedicated follow-up observations at higher spectral and spatial resolution. In fact, we expect that large IFS surveys will provide an essential database allowing the identification of largely undisturbed galaxies (similar to HCG~91c) in the early stage of their interaction with a compact group or cluster. As yet mostly unaffected by their environments, these systems can shine a unique light on the early stages of complex gravitational interactions and the associated consequences.

\section{Summary}\label{sec:summary}

In this article, we presented the discovery of three compact star forming regions with oxygen abundance and kinematic anomalies (at the 1-2 sigma level) in the otherwise unremarkable star forming spiral galaxy HCG~91c. From the analysis of the different strong optical emission lines detected with WiFeS, we found these anomalous star forming regions a) to be comparatively more metal poor by 0.15 dex with respect to their surrounding, and as expected from the overall metallicity gradient present in the inner region of HCG~91c, b) to kinematically lag behind the disk rotation by 5-10 km s$^{-1}$, and c) for one of them, to be associated with the highest value of the ionisation parameter in the entire galaxy ($\log$(q)$\approx$7.95). 

To understand the origin of these peculiar-star forming regions, we combined our WiFeS data set with broad-band images of HCG~91c from Pan-STARRS, and with \emph{VLA} observations of the group-wide {\Hi} distribution of HCG~91. These datasets reveal that HCG~91c is still largely undisturbed, but most likely experiencing the very first stage of its gravitational interactions with other galaxies inside HCG~91 (and possibly with the large tidal tail from HCG~91a). 

Under these circumstances, we discussed three possible scenarios to explain the origin of the anomalous star forming regions detected with WiFeS: accretion of a satellite, gas inflow from the halo, and collapsing pre-existing gas clouds at the disk-halo interface. We found that the latter scenario could naturally explain all of the observed characteristics of the anomalous star forming regions (lower metallicity, velocity lag, compactness, lower reddening). By comparison, the satellite accretion and gas inflows scenarios are harder to reconcile with the observed gas kinematics, but cannot be firmly ruled out at this stage. 

We discussed possible mechanisms able to trigger star formation in these originally stable, pre-existing gas clouds, in the form of tidal shears, long-range gravitational perturbations or harassment from the other group members. Theoretically, these scenarios are consistent with the recent simulations of \cite{Renaud14} suggesting that compressive turbulence is responsible for enhanced star formation activity in the very early stages of galaxy interactions, while large scale gas flows within a galaxy disk feeding central starbursts occur later on at the second closest approach.

HCG~91c may be offering a direct window on the early phase of galaxy interactions in Compact Group environments, and possibly on one of the early stage of galaxy evolution. The existence of localized regions of lower metallicity gas suggest that gas mixing (leading to a flattening of the overall abundance gradient) can be occurring on the scales of individual star forming regions before the onset of large inflows following stronger gravitational effects. In the era of large scale IFS surveys such as CALIFA, SAMI and MaNGA, HCG~91c act as a reminder that mechanisms associated with galaxy evolution may first be impactful on sub-kpc scale and display a discreet kinematic signature ($\Delta$v$\approx$10 km s$^{-1}$ in the present case).

Dedicated follow-up MUSE observations will provide us with a sharper view of HCG~91c and its peculiar star forming regions. This dataset ought to let us better characterise the physical conditions of the ionized gas throughout HCG~91c, refine our detections of lower abundances, kinematic offsets and reduced extragalactic reddening associated with the C1, C2 and C3 star forming regions, as well as provide important information regarding their underlying stellar population. The MUSE observations will be key to test our current favoured scenario for explaining the lower oxygen abundances of the C1, C2 and C3 regions, and let us identify whether additional star forming regions in the outer regions of HCG~91c (undetected with WiFeS) display similar  gaseous abundance anomalies.

\section*{Acknowledgments} 
We thank I-Ting Ho for sharing his \textsc{lzifu} \textsc{idl} routine with us, Bill Roberts and the IT team at the Research School of Astronomy and Astrophysics (RSAA) at the Australian National University (ANU) for their support installing and maintaining the PDF3DReportGen software on the school servers, and the anonymous referee for his/her constructive suggestions. Vogt acknowledges a Fulbright scholarship, and further financial support from the Alex Rodgers Travelling scholarship from the RSAA at the ANU. Vogt is also grateful to the Department of Physics and Astronomy at Johns Hopkins University for hosting him during his Fulbright exchange. Dopita acknowledges the support of the Australian Research Council (ARC) through Discovery project DP130103925, and additional financial support for this project from King Abdulaziz University. This research has made use of the \textsc{aladin} interactive sky atlas \citep{Bonnarel00}, and of NASA's Astrophysics Data System and the NASA/IPAC Extragalactic Database (NED) which is operated by the Jet Propulsion Laboratory, California Institute of Technology, under contract with the National Aeronautics and Space Administration. This research also made use of \textsc{statsmodel} \citep{Seabold10}, of \textsc{matplotlib} \citep{Hunter07}, of \textsc{astropy}, a community-developed core \textsc{python} package for Astronomy \citep{astropy13}, of \textsc{mayavi} \citep{Ram11}, of \textsc{aplpy}, an open-source plotting package for \textsc{python} hosted at http://aplpy.github.com, and of \textsc{montage}, funded by the National Aeronautics and Space Administration's Earth Science Technology Office, Computation Technologies Project, under Cooperative Agreement Number NCC5-626 between NASA and the California Institute of Technology. \textsc{montage} is maintained by the NASA/IPAC Infrared Science Archive. The ``Second Epoch Survey" of the southern sky was made by the Anglo-Australian Observatory (AAO) with the UK Schmidt Telescope. Plates from this survey have been digitized and compressed at the Space Telescope Science Institute under U.S. Government grant NAG W-2166. The Pan-STARRS1 Surveys (PS1) have been made possible through contributions of the Institute for Astronomy, the University of Hawaii, the Pan-STARRS Project Office, the Max-Planck Society and its participating institutes, the Max Planck Institute for Astronomy, Heidelberg and the Max Planck Institute for Extraterrestrial Physics, Garching, The Johns Hopkins University, Durham University, the University of Edinburgh, Queen's University Belfast, the Harvard-Smithsonian Center for Astrophysics, the Las Cumbres Observatory Global Telescope Network Incorporated, the National Central University of Taiwan, the Space Telescope Science Institute, the National Aeronautics and Space Administration under Grant No. NNX08AR22G issued through the Planetary Science Division of the NASA Science Mission Directorate, the National Science Foundation under Grant No. AST-1238877, the University of Maryland, and Eotvos Lorand University (ELTE). We thank the PS1 Builders and PS1 operations staff for construction and operation of the PS1 system and access to the data products provided.

\bibliographystyle{mn2e_new}
\bibliography{vogt_2015}

\newpage
\appendix
\section{Comparison of the gas kinematics with Amram et al. (2003) }\label{app:amram}

\cite{Amram03} presented Fabry-Perot observations of the H$\alpha$ emission line in the galaxy HCG~91c, and reported the existence of large asymmetries and multiple distinct components in the line profile (see their Figure~20). Their data was acquired with the CIGALE Fabry-Perot mounted on the \emph{ESO 3.6m telescope} at La Silla on 1995 August 21-24 with a seeing of $\sim$1 arcsec. The spectral resolution of their dataset was R=9375 at H$\alpha$, with 24 scanning steps and a sampling step of 0.35{\AA} or $\sim$16 km s$^{-1}$, and a spatial pixel size of 0.91$\times$0.91 arcsec$^{2}$. 

\begin{figure*}
\centerline{\includegraphics[scale=0.5]{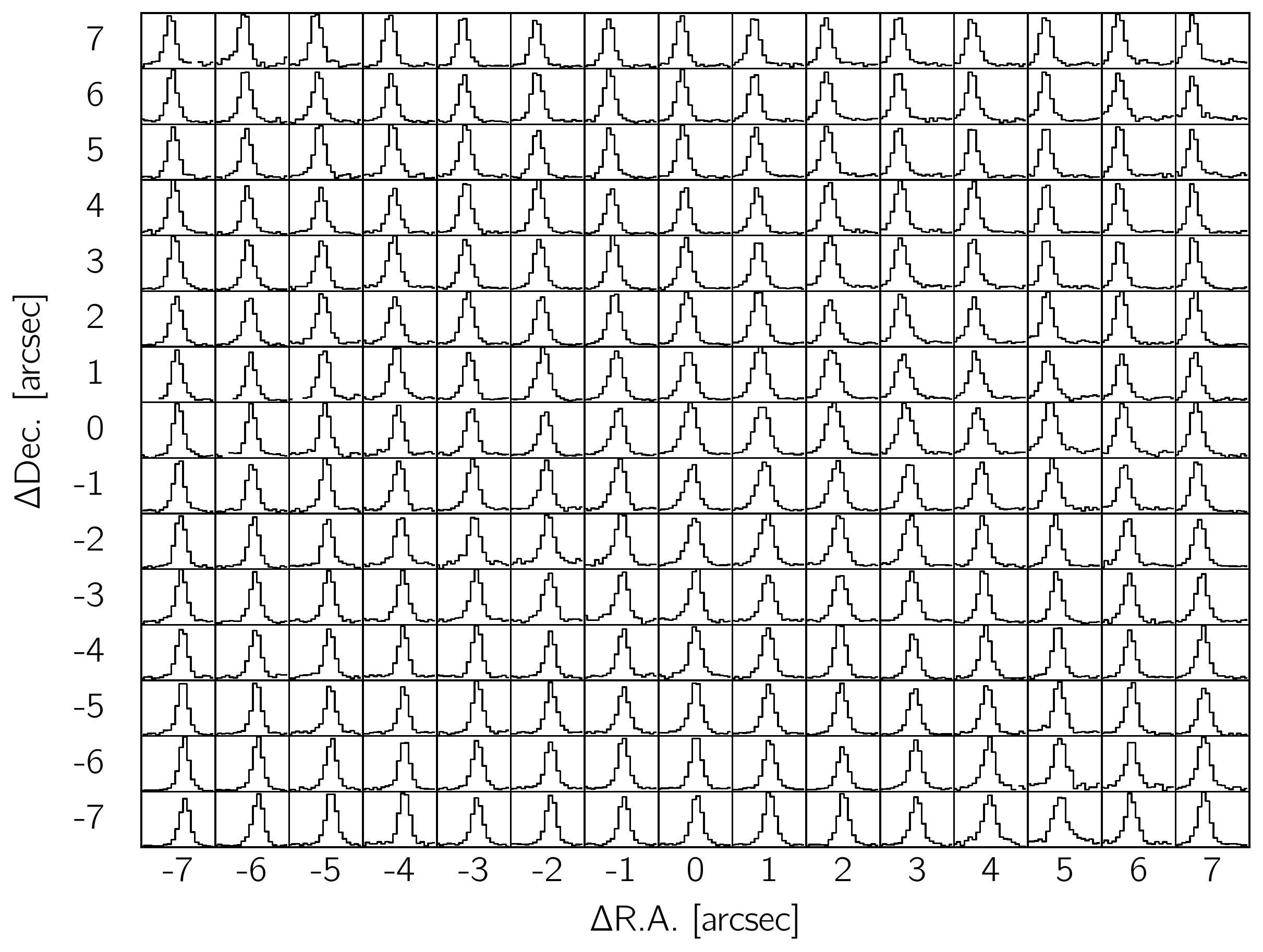}}
\caption{Reproduction of Figure~20 of Amram et al. (2003), but for our WiFeS observations: the H$\alpha$ line profile for all the spaxels within a box of 15$\times$15 arcsec$^2$ centered on the galaxy's core (i.e. spaxel [0;0] in our reference frame). Each panel contains 19 wavelengths bins ranging from 6719.56{\AA} to 6727.48{\AA}. Each panel is normalized to the emission line peak. }\label{fig:amram}
\end{figure*}

In Figure~\ref{fig:amram}, we reproduce the Figure~20 of \cite{Amram03} showing the H$\alpha$ line profile for the central 15$\times$15 arcsec$^2$ regions of HCG~91c, but for our WiFeS data. Modulo minor differences (i.e. each panel corresponds to 1$\times$1 arcsec$^2$ instead of 0.91$\times$0.91 arcsec$^2$), both Figures are directly comparable. Especially, each panel in our Figure~\ref{fig:amram} covers the same spectral range as Figure~20 of \cite{Amram03}. 

These authors reports secondary velocity components redshifted by $\sim$100 km s$^{-1}$ from the main line peak, and with dispersion extending over the entire spectral range shown in the different panels. It is clear from Figure~\ref{fig:amram} that despite the slightly lower spectral resolution (R=7000) and spectral sampling (0.44{\AA} or $\sim$19.5 km s$^{-1}$) of our dataset, WiFeS would have detected complex line profiles such as those reported by \cite{Amram03}. Instead, the H$\alpha$ line profiles observed by WiFeS are narrow (with $\sigma\approx$20-40 km s$^{-1}$, see Section~\ref{sec:wifes_kin}) and single-peaked. We detect some small asymmetries in the inner most spaxels consistent with beam smearing. We note that these asymmetries are mostly \emph{blueshifted} with respect to the main line peak, while the asymmetries reported by \cite{Amram03} are largely \emph{redshifted}.

As we find no evidence for multiple kinematics components in our WiFeS observations of HCG~91c - and especially find no evidence for the complex line profiles reported by \cite{Amram03} - we are lead to conclude that the gas in this galaxy can be well described by a single kinematic structure at all locations.

\section{Error propagation in \textsc{pyqz} v0.6.0}\label{app:pyqz}

In the original version published by \cite{Dopita13a}, the \textsc{pyqz} v0.4.0 module was not set up to account for errors in the flux measurements of emission lines. Instead, errors on the estimation of $\log$(q) and 12+$\log$(O/H) were estimated from the standard deviation between the different estimates stemming from the different line ratio diagnostic grids chosen by the user. Indeed, for flux measurements at the $\sim$5\% level, the mismatch between the estimates from different diagnostics typically dominates the uncertainty of the joint final estimate. For larger observational errors on the emission line fluxes, \textsc{pyqz} v0.4.0 therefore under-estimates the true error associated with the final values of 12+$\log$(O/H) and $\log$(q). Here, we have upgraded \textsc{pyqz} to address this shortcoming. The updated version of \textsc{pyqz} will be publicly released and described in details in the near future along with new \textsc{mappings iv} models. Here, we briefly describe for completeness how errors on the final 12+$\log$(O/H) and $\log$(q) estimates in our analysis of HCG~91c are computed. 

The underlying idea is to propagate the full probability density function associated with each line flux measurement to the 12+$\log$(O/H) \emph{vs} $\log$(q) plane. To that end, for each set of line fluxes (and errors) provided by the user (i.e. for each individual spaxel, in the case of our analysis), \textsc{pyqz} v0.6.0 generates $n$ random set of line fluxes from the errors provided by the user, assuming a normal and uncorrelated error distribution for each line. \textsc{pyqz} subsequently derives an estimate of 12+$\log$(O/H) and $\log$(q) for all $n$ set of random line fluxes, and for every line ratio diagnostic grid selected by the user, resulting in $n\times m$ estimates of 12+$\log$(O/H) and $\log$(q) (where $m$ is the number of line ratio grids selected by the user). Effectively, these $n\times m$ estimates represent the discretized two-dimensional joint probability distribution function of the oxygen abundance and ionization parameter. The full joint probability density function can the be reconstructed by performing a two-dimensional kernel density estimation. This frequentist approach to error estimation differs from the Bayes formalism adopted by \cite{Blanc14} in the \textsc{izi} code written in \textsc{idl} with a similar purpose to \textsc{pyqz}. 

It is often practical to derive from the joint probability density function a \emph{best-estimate} and associated uncertainty. To derive the best estimate for 12+$\log$(O/H) and $\log$(q), \textsc{pyqz} normalizes the reconstructed joint probability function to its peak level, and computes the 61\%-level contour (corresponding the 1-sigma level for a normal distribution). The mean location of this contour is then chosen as the best estimate of the 12+$\log$(O/H) and $\log$(q), while the full extent of the contour is indicative of the uncertainty associated with these estimates. 

\begin{figure*}
\centerline{\includegraphics[scale=0.35]{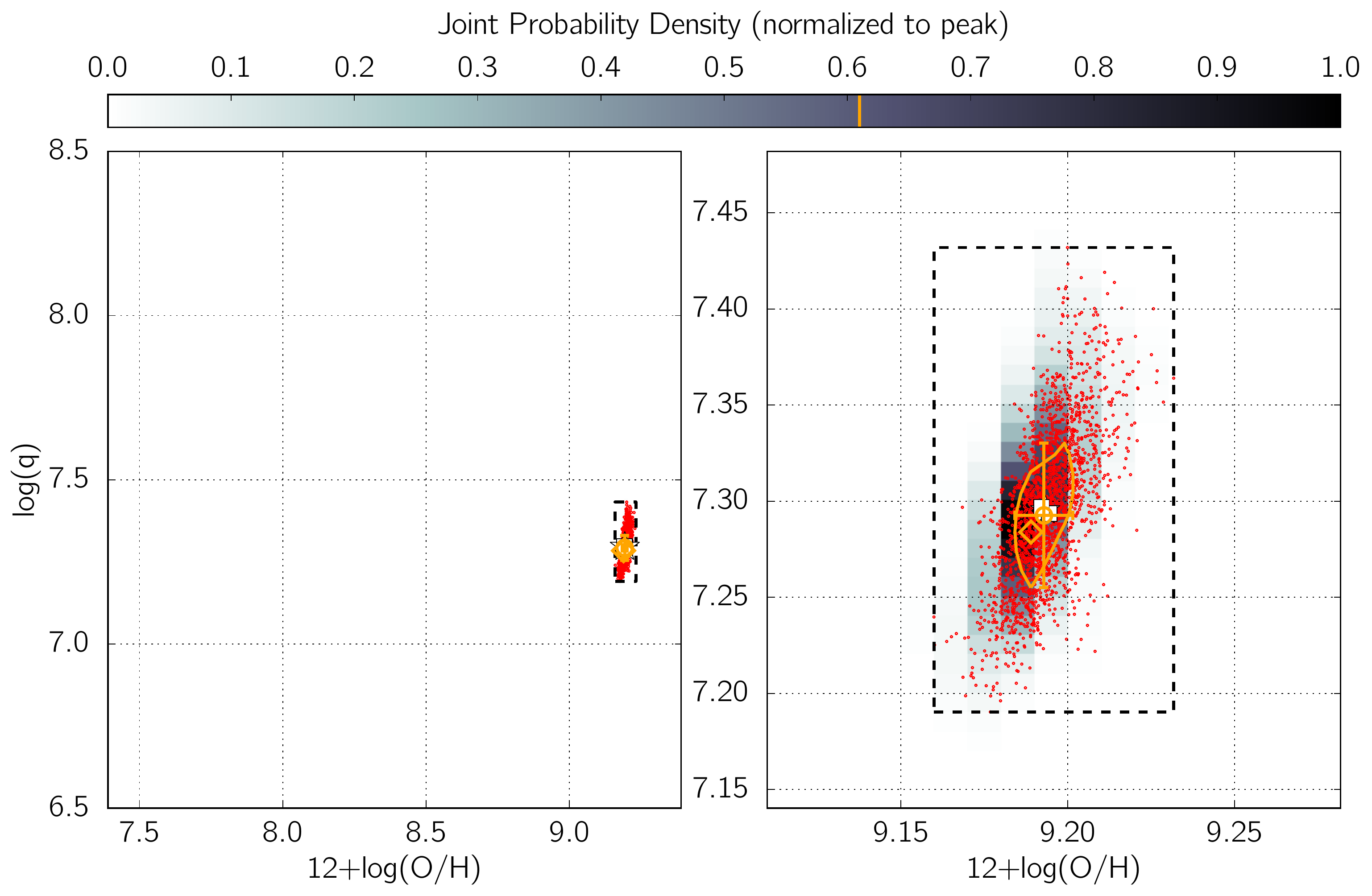}}
\caption{Full (left) and magnified (right) oxygen abundance and ionization parameter plane accessible to \textsc{pyqz} v0.6.0. The data corresponds to the spaxel [0;0] in our observations of HCG~91c. The white squares mark the two estimates associated with the diagnostic grids $\log$~$[${\Nii}$]$/[{\Sii}] \emph{vs} $\log$~[{\Oiii}]/H$\beta$ and $\log$~$[${\Nii}$]$/[{\Sii}] \emph{vs} $\log$~[{\Oiii}]/[{\Sii}]. The red dots mark the location of 2$\times$1000 \emph{randomly-reconstructed} line fluxes from the original line fluxes and associated errors. The reconstructed 2-D joint probability density distribution (with a fixed resolution of 0.01 in 12+$\log$(O/H) and 0.01 in $\log$(q)) is also shown, with the orange contour tracing the 1-sigma level from the peak. The peak location is marked with an orange diamond, while the mean location of the 1-sigma contour and its extent in marked with the orange circle and associated error bars. The dashed box marks the full extent of the random set of individual estimates (the red dots). }\label{fig:pyqz1}
\end{figure*}

\begin{figure*}
\centerline{\includegraphics[scale=0.35]{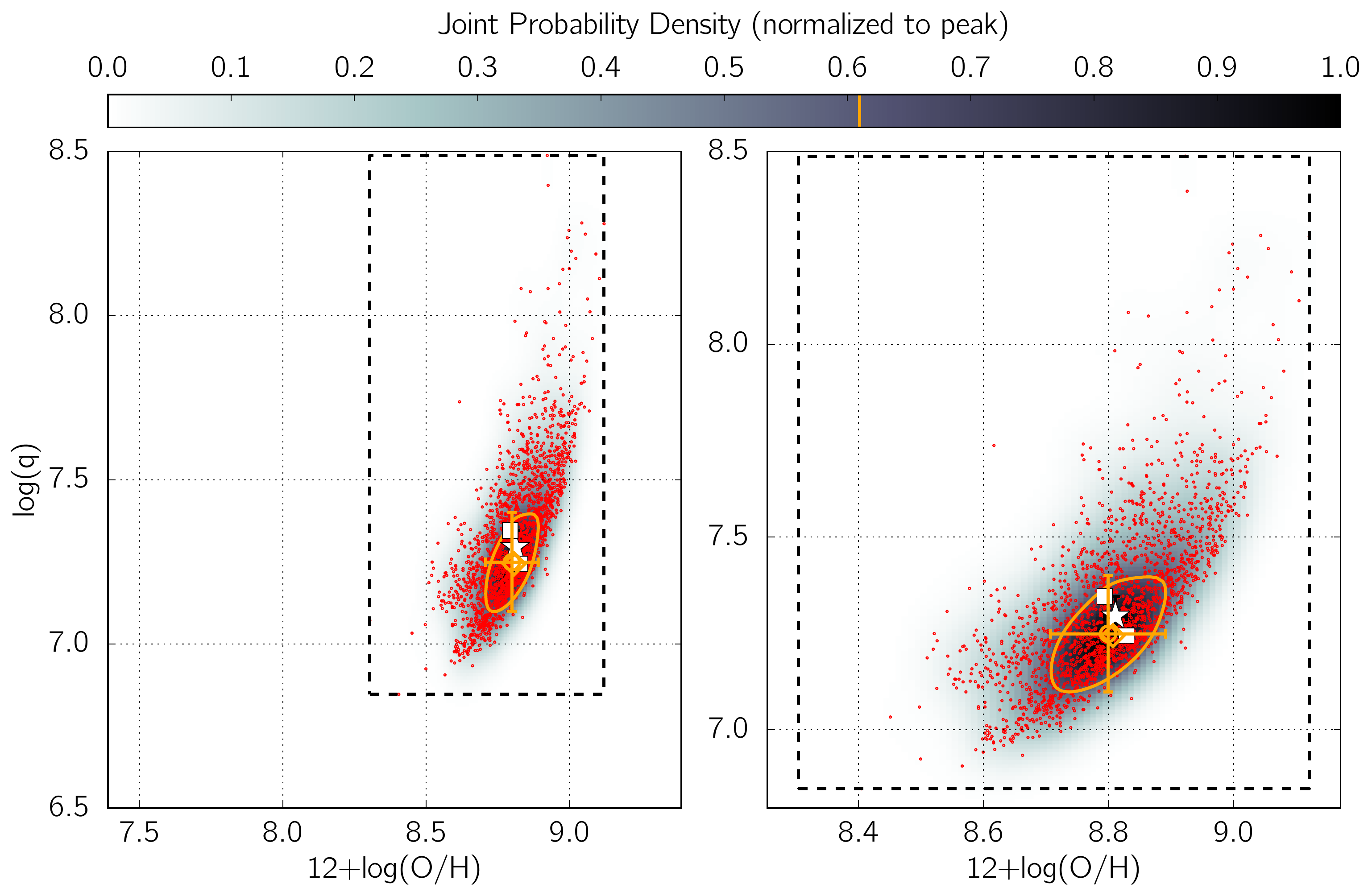}}
\caption{Same as Figure~\ref{fig:pyqz1}, but for the spaxel [9;14]. The lower S/N in the data results in a larger spread of the random set of estimates.}\label{fig:pyqz2}
\end{figure*}

The process is illustrated for two representative high and low S/N spaxels in our WiFeS observations of HCG~91c in Figures~\ref{fig:pyqz1} and \ref{fig:pyqz2}, respectively. For reasons detailed in Section~\ref{sec:pyqz}, these examples rely on only two line ratio diagnostic grids: $\log$~$[${\Nii}$]$/[{\Sii}] \emph{vs} $\log$~[{\Oiii}]/H$\beta$ and $\log$~$[${\Nii}$]$/[{\Sii}] \emph{vs} $\log$~[{\Oiii}]/[{\Sii}]. The location of the individual estimates stemming from either diagnostic grids in the 12+$\log$(O/H) \emph{vs} $\log$(q) are marked with white squares. The mean of these positions, corresponding to the \textsc{pyqz} v0.4.0 estimate of 12+$\log$(O/H) and $\log$(q), is marked with a white star. For both spaxels, 1000 random set of line fluxes were generated by \textsc{pyqz}, resulting in 2000 estimates of 12+$\log$(O/H) and $\log$(q) (1000 for either line ratio grid), shown as individual red dots. 

The 61\%-of-peak-level contour of the underlying reconstructed joint probability density function is shown with an orange line. With 1000 random sets of line fluxes, we find that the location of the peak of the joint probability density function (marked with an orange diamond symbol) varies within the 1-sigma contours when \textsc{pyqz} is run multiple times. However, the location of the 1-sigma contour at 61\% of the peak level remains consistent when \textsc{pyqz} is run multiple times. This fact motivates our choice to base the final estimates of 12+$\log$(O/H) and $\log$(q) on the mean location of the 1-sigma contour level, rather than the exact peak location. 

Different \textsc{python} tools exists to perform a two-dimensional kernel density estimation. \textsc{pyqz} can be set by the user to use two of these - either the \textsc{gaussian\_kde} routine in the \textsc{scipy.stats} package \footnote{http://docs.scipy.org/doc/scipy-0.14.0/reference/generated/ scipy.stats.gaussian\_kde.html, accessed on 2014 December 13.}, or the \textsc{kdemultivariate} routine in the \textsc{statsmodel} package \footnote{http://statsmodels.sourceforge.net/devel/generated/ statsmodels.nonparametric.kernel\_density.KDEMultivariate.html, accessed on 2014 December 13}. The former routine is 10-100 times faster but is not well suited for bi- or multi-modal distributions. In the case of \textsc{pyqz}, the \textsc{gaussian\_kde} routine tends to over-smooth the joint probability function when two diagnostics are not directly consistent with one another (for example in the case of a flux measurement error). The option to use the \textsc{gaussian\_kde} routine is included in \textsc{pyqz} v0.6.0 as a rapid alternative to test the code and obtain preliminary results. In practice, it is preferable to use the \textsc{kdemultivariate} routine instead which provides more accurate results, as the bandwidth of the gaussian kernel can be set individually for the 12+$\log$(O/H) and $\log$(q) directions. All estimates of oxygen abundances and ionization parameters presented in this article were derived using the \textsc{kdemultivariate} method.

\label{lastpage}
\end{document}